\documentclass[twoside]{elsart} 
\journal{Astroparticle Physics}
\usepackage[dvips]{graphicx}
\usepackage[dvips]{color}
\usepackage{natbib}
\usepackage{amssymb}  

\begin{document}
\def\en{E_{\nu}}
\def\eg{E_{\gamma}}
\def\ep{E_{p}}
\def\epb{\epsilon_{p}^{b}}
\def\enb{\epsilon_{\nu}^{b}}
\def\enbG{\epsilon_{\nu,GeV}^{b}}
\def\enbM{\epsilon_{\nu,MeV}^{b}}
\def\ens{\epsilon_{\nu}^{s}}
\def\ensG{\epsilon_{\nu,GeV}^{s}}
\def\egb{\epsilon_{\gamma}^{b}}
\def\egbM{\epsilon_{\gamma,MeV}^{b}}
\def\g25{\Gamma_{2.5}}
\def\lumi{L_{\gamma}^{52}}
\begin{frontmatter}
\title{Coincident GRB neutrino flux predictions:\\{\Large Implications for experimental UHE neutrino physics}}
\author[dort]{Julia K.~Becker\corauthref{cor}}
\author[mad]{, Michael Stamatikos}
\author[mad]{, Francis Halzen}
\author[dort]{, Wolfgang Rhode}
\corauth[cor]{{\scriptsize Corresponding author. Contact: julia@physik.uni-dortmund.de, phone: +49-231-7553667}}
\address[dort]{Department of Physics, Dortmund University, D-44221 Dortmund, Germany}
\address[mad]{Department of Physics, University of Wisconsin, Madison, WI-53706}
\date{\today}
\begin{abstract}
In the hadronic fireball phenomenology of Gamma Ray Bursts (GRBs), it is expected that the observed photons are accompanied by 
UHE neutrinos, which have not been observed yet. It is one of the challenges of experimental UHE neutrino astrophysics 
to look for a signal from GRBs. In this paper, the differences between a search for a diffuse signal and an examination 
of a source sample given by e.g.~BATSE will be analyzed. Since redshift
information is needed to determine the correct energy spectrum, long duration
bursts with redshifts from different estimate methods will be used. We will start with an overview of the current understanding of 
GRB neutrino physics and will then use this knowledge to make predictions for a coincidence flux and a corresponding diffuse flux.
It can be shown that shape and normalization of the spectrum is highly dependent on the set
of bursts used and that individual bursts can determine
the total spectrum.
\end{abstract}
\begin{keyword}
Neutrinos \sep GRBs \sep lag \sep variability \sep single source predictions
\PACS 95.55.Vj \sep 96.60.Rd \sep 98.62.Nx \sep 98.38.Fs
\end{keyword}
\end{frontmatter}
\section{Introduction \label{introduction}}
The prompt GRB photon spectrum, $N_{\gamma}$, is usually given by a Band function~\cite{band} which can be approximated
by a broken power law,
\begin{equation}
N_{\gamma}\propto \left\{ \begin{array}{lll}
\eg^{\alpha_{\gamma}}&&\mbox{for } \eg<\egb\\
\eg^{\beta_{\gamma}}&&\mbox{for } \eg\geq \egb\,.
\end{array}
\right.
\end{equation}
The spectrum is presumably produced by synchrotron radiation of electrons in
the internal
shock fronts of the jet, see e.g.\ \cite{hh_02}. There are two approaches to
explain the break in the 
spectrum at a break energy of typically $\egb\sim 250$~keV: The most common
explanation is the steepening of the spectrum by a power of one due to cooling
of electrons at high energies, see for example~\cite{zhang,piran} as a
review. The break can, however, also be explained by assuming an Inverse
Compton scattering scenario, see e.g.~\cite{dar_rujula} and references therein. Throughout the paper, all given energies are in the
observer's frame at Earth unless declared otherwise. 
The spectral indices are usually scattered around average values of $\alpha_{\gamma}\sim -1$ and $\beta_{\gamma}\sim -2$.
Assuming hadronic acceleration in the jet, a prompt neutrino flux that is correlated
to the photon spectrum results from photohadronic interactions
in the source. The neutrino spectrum, $dN_{\nu}/d\en$, can be derived assuming
that the proton spectrum follows the electron spectrum of the source. Since
the neutrino flux in turn follows the proton spectrum in first order
approximation, it can thus be connected to the observed synchrotron spectrum
of the sources and can be described as
\begin{equation}
\frac{dN_{\nu}}{d\en}\,\en^{2}=A_{\nu}\cdot \left\{ \begin{array}{lll}
(\en/\enb)^{-\alpha_{\nu}}&&\mbox{for } \en<\enb\\
(\en/\enb)^{-\beta_{\nu}}&&\mbox{for } \enb<\en\leq \ens\\
(\en/\enb)^{-\beta_{\nu}}\,(\en/\ens)^{-2}&&\en\geq \ens\,.
\end{array}
\right.
\end{equation}
The photon spectral indices can be used to describe $\alpha_{\nu}=\beta_{\gamma}+1$ and
$\beta_{\nu}=\alpha_{\gamma}+1$. 
The second break at $\en=\ens$ in the neutrino spectrum results from the fact that pions lose energy at very high energy due to synchrotron radiation.
Thus, less neutrinos result from pions at very high energies which leads to a steepening of the spectrum by a power of 2. A detailed
derivation of the neutrino spectrum as presented above is given in
\cite{wax_equi}. 

The spectrum is normalized to the $\gamma$-ray fluence $F_{\gamma}$ which is
assumed to be proportional to the neutrino luminosity,
\begin{equation}
x\cdot
F_{\gamma}=\int_{E_{\min}}^{E_{\max}}\frac{dN_{\nu}}{dE_{\nu}}\,dE_{\nu}\approx
\ln(10)\cdot A_{\nu}\,.
\end{equation}
 All parameters occurring in following calculations are listed in table \ref{parameters}.
The factor $x$ is given by the product of the fraction of proton energy
transfered to the pions, $f_{\pi}$, a factor $1/8$ since half of the
photohadronic interactions result in four neutrinos and a factor
$1/f_e$ to account for the fraction of total energy in electrons compared to protons in the jet~\cite{guetta}. The normalization constant $A_{\nu}$ is therefore given
as
\begin{equation}
A_{\nu}=\frac{1}{8}\frac{1}{f_e}\frac{F_{\gamma}}{\ln(10)}f_{\pi}\,.
\end{equation}
In the following, the normalization of a single burst will be modified to a
quasi-diffuse normalization by multiplying $A_{\nu}$ with the number of bursts
per year ($2/3\cdot 1000$ long duration bursts per year) and dividing the
result by $4\,\pi$ sr,
\begin{equation}
A_{\nu}'=\frac{2}{3}\cdot \frac{1000}{\mbox{yr }4\,\pi\mbox{ sr}}A_{\nu}\,.
\end{equation}

The first break energy in the spectrum, $\enb$, is related to the break energy in the photon spectrum as
\begin{equation}
\enb= \frac{(m_{\Delta}^{2}-m_{p}^{2})\cdot \Gamma^{2}}{4\cdot (1+z)^{2}}\cdot\left(\egb\right)^{-1}\mbox{ GeV}\,.
\end{equation}
It is determined through the minimal energy necessary to produce a
$\Delta$-resonance in the shock fronts of the bursts. 
Using the numerical values given in~\cite{ppb} for the proton mass, $m_{p}=0.94$~GeV,
and the $\Delta$ mass, $m_{\Delta}=1.23$~GeV,  leads to 
\begin{equation}
\enb=7\cdot10^{5}\cdot (1+z)^{-2}\,\frac{\g25^{2}}{\egbM}\mbox{ GeV}\,.
\label{enb_equ}
\end{equation}

The second break energy is connected to the synchrotron loss time. It depends on the neutrino flavor and for muon neutrinos, it 
is given as
\begin{equation}
\ens=\sqrt{\frac{3\,\pi\,\epsilon_e}{4\,\tau_{\pi}^{0}\,\sigma_{T}\,\epsilon_B\,L_{\gamma}}}\cdot
\frac{c^4\,t_v}{(1+z)\cdot m_{e}}\,\Gamma^{4}\,.
\label{ens_1}
\end{equation}

For electron and anti-muon neutrinos the break energy $\ens$ is about an order of
magnitude lower, since these neutrinos result from the muon decay. The muon
lifetime is about a factor of 100 higher than the pion lifetime, which reduces
the threshold of synchrotron losses.
The derivation of the second break energy can be found in~\cite{guetta}. Here,
$\epsilon_B$ and $\epsilon_e$ are the fraction of the burst's internal energy going into the
magnetic field respectively into electrons. 
The equipartition fractions have been set to $\epsilon_e=0.1$ and
$\epsilon_b=0.1$. There is no good way of determining the equipartition
fractions theoretically yet. However, afterglow observations indicate values of
the order of $0.1$ \cite{wax_equi}. The remaining parameters in
Equ.~(\ref{ens_1}) are listed in table~\ref{parameters} with the values as used in the
following calculations. Inserting all numerical values gives
\begin{equation}
\ens=\frac{10^{8}}{1+z} \epsilon_{e}^{1/2} \, \epsilon_{b}^{-1/2}\, \g25^{4} \,t_{v,-2}/\sqrt{\lumi} \mbox{GeV}\,.
\label{ens_equ}
\end{equation}

The boost factor $\Gamma$ is constrained to $100<\Gamma<1000$, since
for boost factors less than $100$, the medium would be optically thick to photons and
for $\Gamma>1000$, since protons lose most of their energy to synchrotron radiation. The possibility of fluctuating
$\Gamma$ using the photon break energy is given as demonstrated in \cite{guetta}, but there are several arguments for
using a constant value: Bursts can be misaligned which would lead to a
misinterpretation of the boost factor. Also, varying the break energy for each
single burst might implicitly already include boost factor fluctuations. Thus,
a constant boost factor of $\Gamma=300$ is used in following calculations.

In previous papers, e.g.~\cite{guetta}, a variation of the fraction of energy
going into pions $f_{\pi}$ has been discussed. Such a variation would further increase
the width of the distribution of the neutrino spectrum normalization. $f_{\pi}$ varies with the burst
  luminosity, the boost factor, the photon break energy and the variability time as
\begin{equation}
f_{\pi}\sim 0.2\cdot \frac{\lumi}{\g25^{4}\,t_{v,-2}\,\egbM}\,.
\label{fpi}
\end{equation}
In the following, this fraction will be kept at a constant value of
$f_{\pi}\sim0.2$ for the following reasons: 
\begin{itemize}
\item $f_{\pi}$ strongly depends on the
boost factor $\Gamma$ which will be used as a constant as discussed
before. The dependence on the other three parameters is only linearly and thus
not as striking as a variation of $\Gamma$ would be. 
\item The main uncertainties in the current calculations result from the lack
  of knowledge of the parameters $\epsilon_e$ and $\epsilon_B$ as well as from
  uncertainties in the redshift relation which leads to uncertainties in
  $L_{\gamma}$. These three parameters are all important in the determination
  of the spectrum normalization and thus, a constant value is favorable.
\end{itemize}

The variability time appears in Equ.~(\ref{ens_equ}) as well. Apart from the
arguments mentioned for the normalization of the spectrum already, the
effect of varying $t_{v}$ here would not be very significant in the detection
rates, because the dominant contribution seen in the detector comes from the
energy region of the first break.

The diffuse neutrino flux prediction derived from the model above is given in
\cite{WB}, where the authors use average parameters to determine the shape
of the spectrum. Cosmological evolution of the sources have been considered in
that model by taking into account the redshift evolution of GRBs. It is
assumed that GRBs follow Star Formation Rate, since they appear to be
connected to Supernova-Ic explosions. The final result of this work will be compared
to this standard flux. To do this, the diffuse Waxman-Bahcall flux has to be
weighted by a factor $2/3$, since this work only deals with long duration
bursts which make up approximately $2/3$ of the total burst population.

A different approach to predicting the neutrino flux from GRBs is to look at each burst individually and add up the individual
spectra to make a prediction of the total flux from these sources. An approach
of using individual spectra as discussed by Guetta et al.\ \cite{guetta} will be used here to
get an estimate of the flux that can be expected to be observed by coincident
measurements of the AMANDA\footnote{Antarctic Muon And Neutrino Detector
  Array} telescope. 
In addition, 
a diffuse prediction will be made, using the mean values of the parameter distributions that are given or
can be derived from BATSE data. 

Two different burst samples based on the determination of burst redshifts using the redshift estimators
variability (568 bursts) and lag (292 bursts) will be examined in this
paper. A more detailed discussion of the sample properties will be discussed in section~\ref{samples}.

\begin{table}[ht]
\centering{ \small 
\begin{tabular}{l|l|l}
\hline 
Parameter&Symbol&Typical value\\ \hline \hline
Observed photon flux&$F_{\gamma}$&$\sim 10^{-4}-10^{-6}$~erg/s\\ \hline
Redshift&$z$& $1-2$ \\ \hline
Luminosity distance&$d_l$&$\sim 6.5-15$~Gpc\\ \hline
Photon energy&$\eg$&-\\ \hline
Neutrino energy&$\en$&-\\\hline
Equipartition fractions&$\epsilon_b$&$\sim 0.1$\\
&$\epsilon_e$&$\sim 0.1$\\ \hline
Electron-proton total energy ratio&$f_e$&$\sim 0.1$\\ \hline
Energy fraction transfered from $p$ to $\pi$&$f_{\pi}$&0.2\\ \hline
Burst luminosity&$L_{\gamma}$& $10^{52}$~erg/s\\ 
&$\lumi:=L_{\gamma}/(10^{52}$~erg/s)&1\\ \hline
Boost factor&$\Gamma$& $\sim 300$\\ \hline
&$\Gamma_{2.5}$&$\sim 1$\\ \hline
Spectral indices&$\alpha_{\nu}$& $\sim -1$\\
&$\beta_{\nu}$&$\sim 0$\\ \hline
Photon break energy& $\egb$&$\sim 100-1000$~keV\\
&$\egbM:=\egb/$MeV&$\sim 0.1-1$\\ \hline
Neutrino break energies&$\enb$&$\sim (10^{5}-10^{6})$~GeV\\ 
&$\enbG:=\enb/$GeV&$\sim 10^{5}-10^{6}$\\\hline
&$\ens$&$\sim 10^{7}$~GeV\\\hline
&$\ensG=\ens/$GeV&$\sim 10^{7}$\\\hline
Opening angle&$\theta$&-\\ \hline
Time variability&$t_v$&$\sim 10^{-3}-1$~s\\ 
&$t_{v,-2}:=t_{v}/(10^{-2}$~s)&$\sim 0.1-100$\\ \hline
Burst duration&$t_{90}$&$\sim 2-1000$~s\\ \hline
Pion mass&$m_{\pi}$&$140$~MeV\\\hline
Thompson cross section&$\sigma_{T}$&$0.665\cdot 10^{-24}$~cm$^{2}$\\\hline
pion lifetime at rest&$\tau_{\pi}^{0}$&$2.6\cdot 10^{-8}$~s\\
\end{tabular}
\caption{Parameters for neutrinos flux calculations. The numbers quoted as
  ''typical values'' are to be taken as rough bench marks, since all of these
  parameters fluctuate strongly as emphasized in the text. The calculation of
  the luminosity distance at a redshift between $z=1-2$ is done using
  cosmological parameters of $\Omega_m=0.3$, $\Omega_{\Lambda}=0.7$ and $h=0.71$.}
\label{parameters}
}
\end{table}
\section{The burst samples \label{samples}}
\begin{figure}
\centering{
\includegraphics[width=10cm]{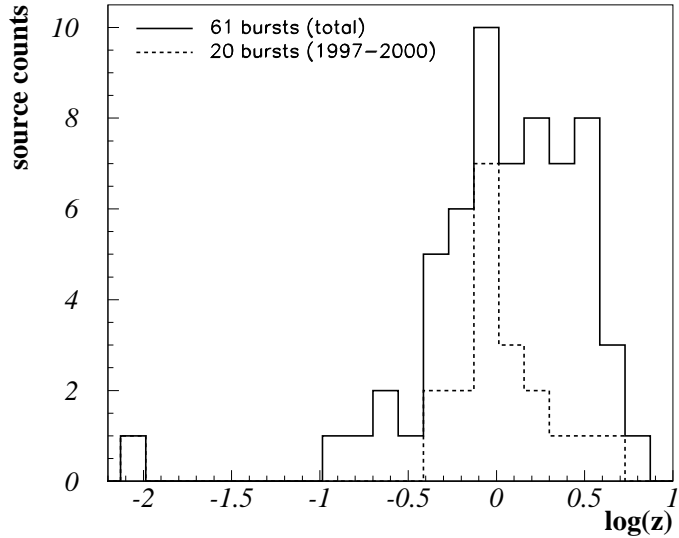}
\caption{Redshift distribution of 61 GRBs with measured redshifts (1997-Oct
  2005, solid line). During the operation of BATSE ($<$June 2000), 20 bursts
  (dashed line) have been assigned redshifts.}
\label{obs_z_distribution}
}
\end{figure}
Due to the necessity of detailed optical afterglow observations for the determination of GRB redshifts, 
there are only about 61 bursts (by Oct 2005) with observed redshifts today. 
A histogram of the 61 observed redshifts 
is presented in Fig.~\ref{obs_z_distribution}. All 61 redshifts observed until
today are shown as the solid line, the dashed line represents the sample of
GRBs with redshift until June 2000, the BATSE era.  
Because of the low statistics of bursts with directly observed redshifts, 
different redshift estimators have been developed. Note that these
methods apply only to \emph{long duration bursts}. Although first measurements of a
short duration burst redshifts succeeded recently, see e.g.~\cite{050509b,050709,050724}, 
there is not enough statistics for short bursts yet. Two burst
samples will be used in the following using different estimator methods for
the redshift determination. Both burst sets are subsamples of the BATSE catalog:

A subsample of 568 bursts with redshifts determined using the variability of
GRBs to estimate redshifts
is given by Guetta et al.\ \cite{guetta}. Throughout the paper, it will be
referred to the \emph{variability sample} when this set of bursts is
discussed. Additionally, a sample of 292 bursts with redshifts determined
connecting temporal lag between the signal in BATSE's different energy
channels is given by \cite{band_lag}, based on the work \cite{norris} and
\cite{bonnell}. Band fits have been applied to fit these spectra. We will refer to this set of GRBs as the
\emph{lag sample}.

The possibility of analyzing bursts with respect to their neutrino signal is
only given for a small set of bursts because of constraints from the
running UHE neutrino experiments. During BATSE's operation time, the AMANDA
experiment is one of the most interesting detectors for GRB analysis: There
are 105 bursts
in the BATSE catalog that can potentially be examined using the AMANDA
experiment now integrated into the IceCube detector as it is pointed out
in~\cite{mike}.  
None of these bursts is given in the lag sample. There are however
82 bursts of the variability sample which are in the set of the 105 bursts as
well. These bursts will be discussed separately in this section. In the
following, we will refer to these bursts as the \emph{AMANDA subsample} for
simplicity. A work particularly dedicated to these about 100 of these bursts will be
presented soon in the context of the analysis of a potential neutrino signal
from these bursts~\cite{mike_in_prep}. Not all of the 105 bursts can actually
be used for an analysis due to the restricted availability of data on either
AMANDA's or Batse's side. A pioneer analysis of a potential neutrino signal from
monster burst GRB030329 in AMANDA is described in~\cite{mike_icrc05}. 

In this section, the parameter distributions will be discussed. Each
individual set of bursts will be used in section \ref{coincidence} to make a
prediction of the flux from the given source sample. Additionally, average
values will be determined by calculating the weighted mean value of each distribution as
\begin{equation}
\overline{p}=\frac{\sum_{i=1}^{n} w(p_i)\cdot p_i}{\sum_{i=1}^{n} w(p_i)}
\end{equation}
with $1\sigma$ errors,
\begin{equation}
\sigma(p)=(\overline{p})^2-\overline{p^2}
\end{equation}
Here, $p_i$ is the corresponding parameter value of a certain bin $i$ and
$w(p_i)$ is the number of entries in that bin. $n$ is the total number of
bins.
In each of the following parameter distributions, the average values are
indicated as vertical lines. The horizontal lines show the $1\sigma$
deviation from the average. The set of average parameters of each sample are
summarized in table~\ref{average_params}.

It should be noted that fit parameters as the spectral indices of the photon
spectra, $\alpha_{\gamma}$ and $\beta_{\gamma}$ as well as the break energy in
the photon spectrum, $\eg$, are not identical for identical bursts in the
different samples. The reason is here that the three parameters are
correlated. Fixing the break energy for example at a certain energy would
influence the values of the spectral indices etc.
\subsection{Lag and Variability samples}
\begin{table}
\begin{tabular}{|c|c|c|c|c|c|}\hline
&$\log\left[A_{\nu}'\right.$/(GeV
    s$^{-1}$sr$^{-1}$cm$^{-2}$)$\left.\right]$&$\alpha_{\nu}$&$\beta_{\nu}$&$\log\enbG$&$\log\ensG$\\
    \hline\hline
variability&$-8.8\pm0.6$&$-0.89\pm0.44$&$-0.17\pm0.37$&$5.0\pm0.8$&$7.0\pm0.9$\\\hline
lag&$-9.0\pm0.7$&$-1.62\pm0.56$&$0.63\pm0.62$&$5.4\pm 0.5$&$7.3\pm0.7$\\\hline
amanda sub.&$-8.9\pm0.5$&$-0.93\pm0.45$&$-0.24\pm0.37$&$4.8\pm0.8$&$6.8\pm1.0$\\ \hline
WB&-8.7&-1&0&5&7\\ \hline
\end{tabular}
\caption{Mean neutrino spectra parameters for the three samples, variability
  (568 bursts), lag (292) and the variability subsample (82 bursts). The
  standard deviation to the mean values has been calculated as an error
  estimate. The values used by Waxman-Bahcall (WB) are given as reference values.}
\label{average_params}
\end{table}

\begin{figure}
\setlength{\unitlength}{1cm}
\begin{minipage}[t][9.5cm][b]{6.7cm}
\begin{picture}(6.7,3.5)
\includegraphics[width=8.cm]{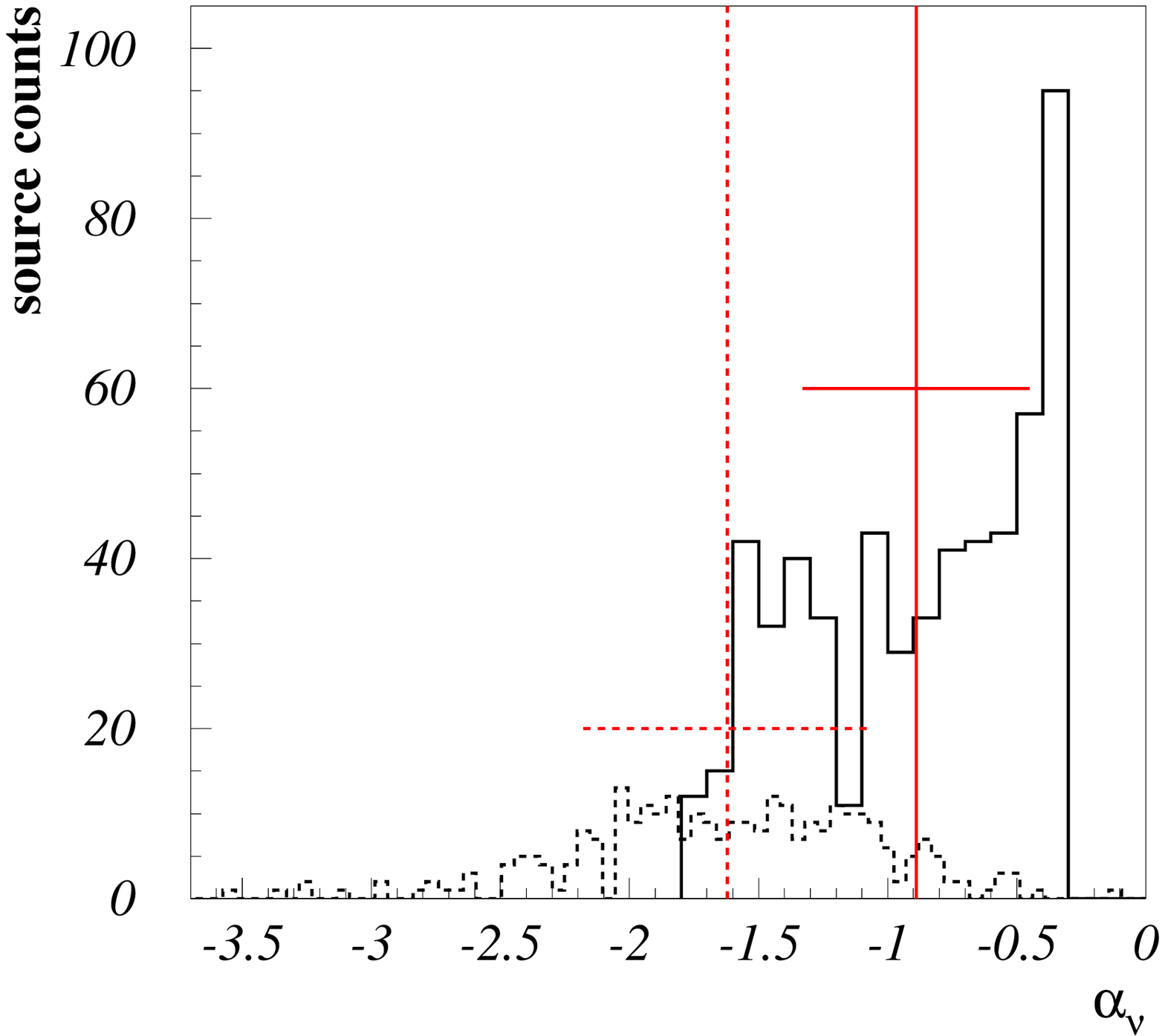}
\end{picture}\par
\caption[]{Distribution of the first spectral index $\alpha_{\nu}$. Solid
  line: variability sample; Dashed line: lag sample. Mean values with
  $1\sigma$ errors are indicated for each distribution.\\[-0.5cm]}
\label{alpha_nu}
\end{minipage}
\parbox{0.5cm}{\quad}
\begin{minipage}[t][9.5cm][b]{6.7cm}
\begin{picture}(6.7,3.5)
\includegraphics[width=8.cm]{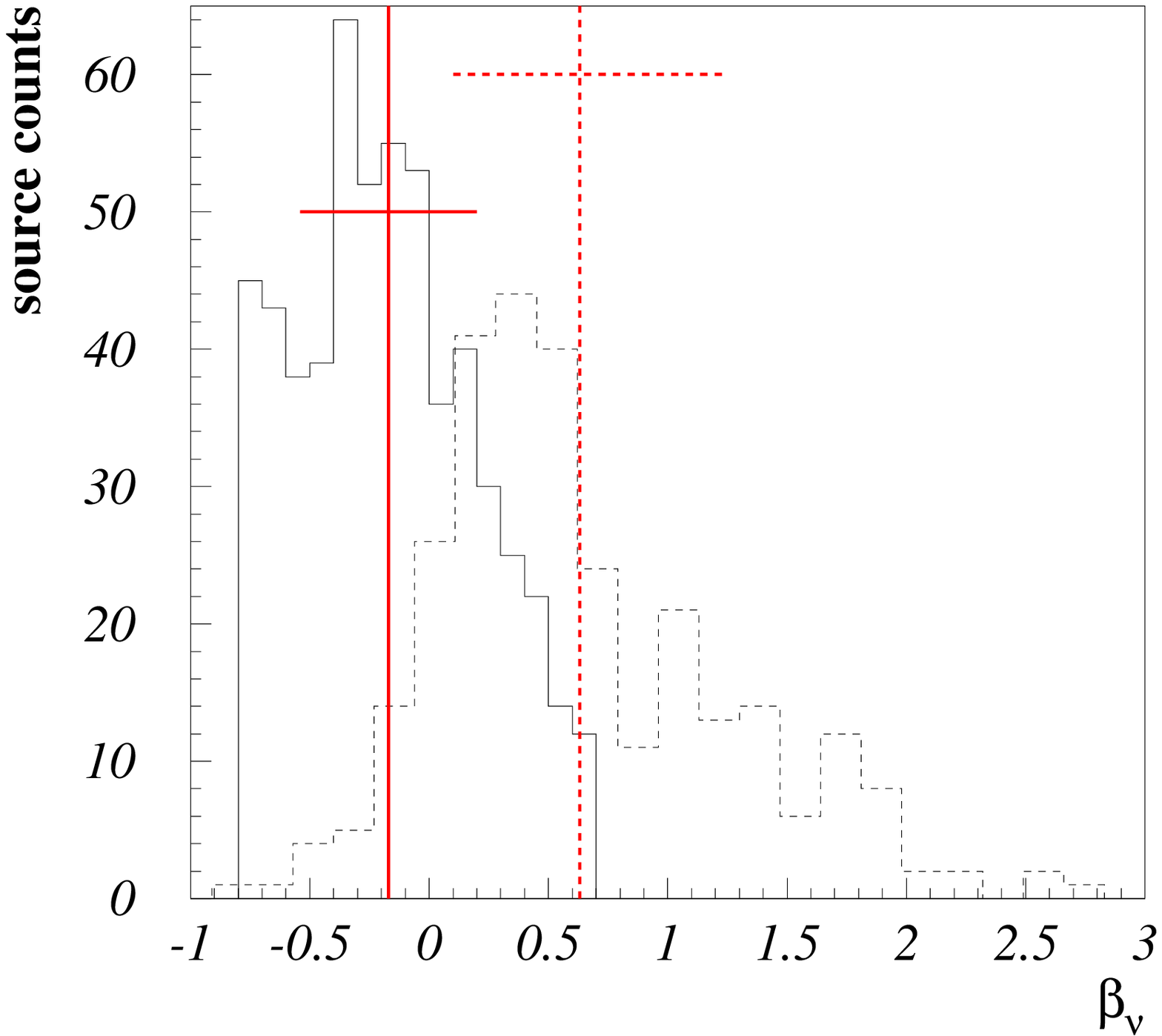}
\end{picture}\par
\caption{Second spectral index, $\beta_{\nu}$, of the neutrino spectrum for bursts in the variability (solid line) and lag sample (dashed line). Mean values with
  $1\sigma$ errors are indicated for each distribution.}
\label{beta_nu}
\end{minipage}\hfill
\end{figure}

The expected neutrino spectrum from individual bursts can be estimated from the parameters of the photon spectrum as
it has been discussed in section \ref{introduction}. 
The distribution of the neutrino spectra's indices follow
the one of the photon spectrum with $\alpha_{\nu}=\beta_{\gamma}+1$ and
$\beta_{\nu}=\alpha_{\gamma}+1$. The distributions of the
spectral indices $\alpha_{\nu}$ and $\beta_{\nu}$ are displayed in Fig.~\ref{alpha_nu} and \ref{beta_nu}. In all following figures, the solid line represents
the distribution of the variability sample, while the dashed line shows the
lag sample distribution. The distributions of the first spectral index differs
between the two samples: While the lag sample shows a Gaussian behavior, the
variability sample distribution does not seem to give a particular
pattern. The distribution seems to be more randomly scattered. The mean values
for the first spectral index are $\overline{\alpha}_{\nu}^{var}=-0.89\pm0.44$ and
$\overline{\alpha}_{\nu}^{lag}=-1.62\pm0.56$. The distributions of the second spectral index also
show different mean values for the two samples, i.e.\
$\overline{\beta}_{\nu}^{var}=-0.17\pm 0.37$ and $\overline{\beta}_{\nu}^{lag}=0.63\pm0.62$.

\begin{figure}[ht]
\centering{
\includegraphics[width=10cm]{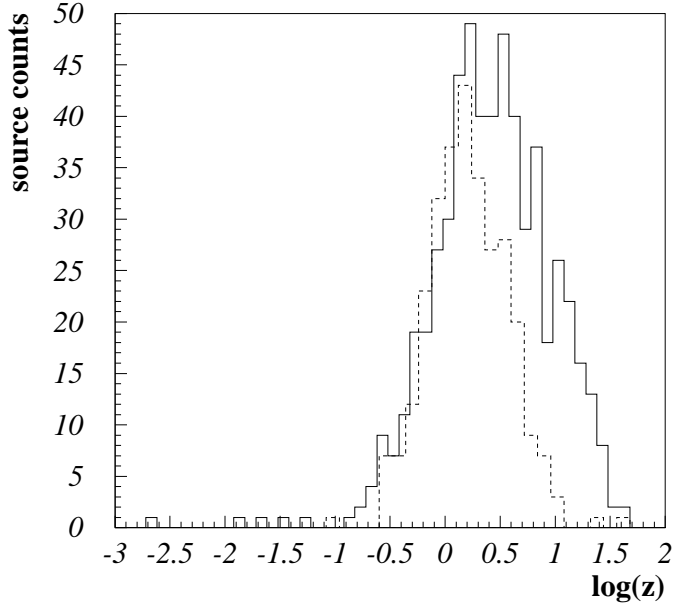}
\caption{Redshift distribution of the sources, solid line: variability sample; dashed
  line: lag sample.}
\label{logz}
}
\end{figure}

The neutrino break energies depend on the luminosity, redshift and the boost
factor of the burst. The redshift distribution is displayed in Fig.~\ref{logz}. The variability sample has a significant contribution in the distribution at higher
redshifts ($0.8<\log z<1.5$), compared to the lag sample. Both distributions have their maxima around $\log z\sim 0-1$ which is consistent
with the recent cognition that GRBs seem to follow Star Formation Rate
(SFR). The boost factor is set to a constant value of $\Gamma=300$ as it has
been discussed above.

\begin{figure}
\setlength{\unitlength}{1cm}
\begin{minipage}[t][9.5cm][b]{6.7cm}
\begin{picture}(6.7,3.5)
\includegraphics[width=8.cm]{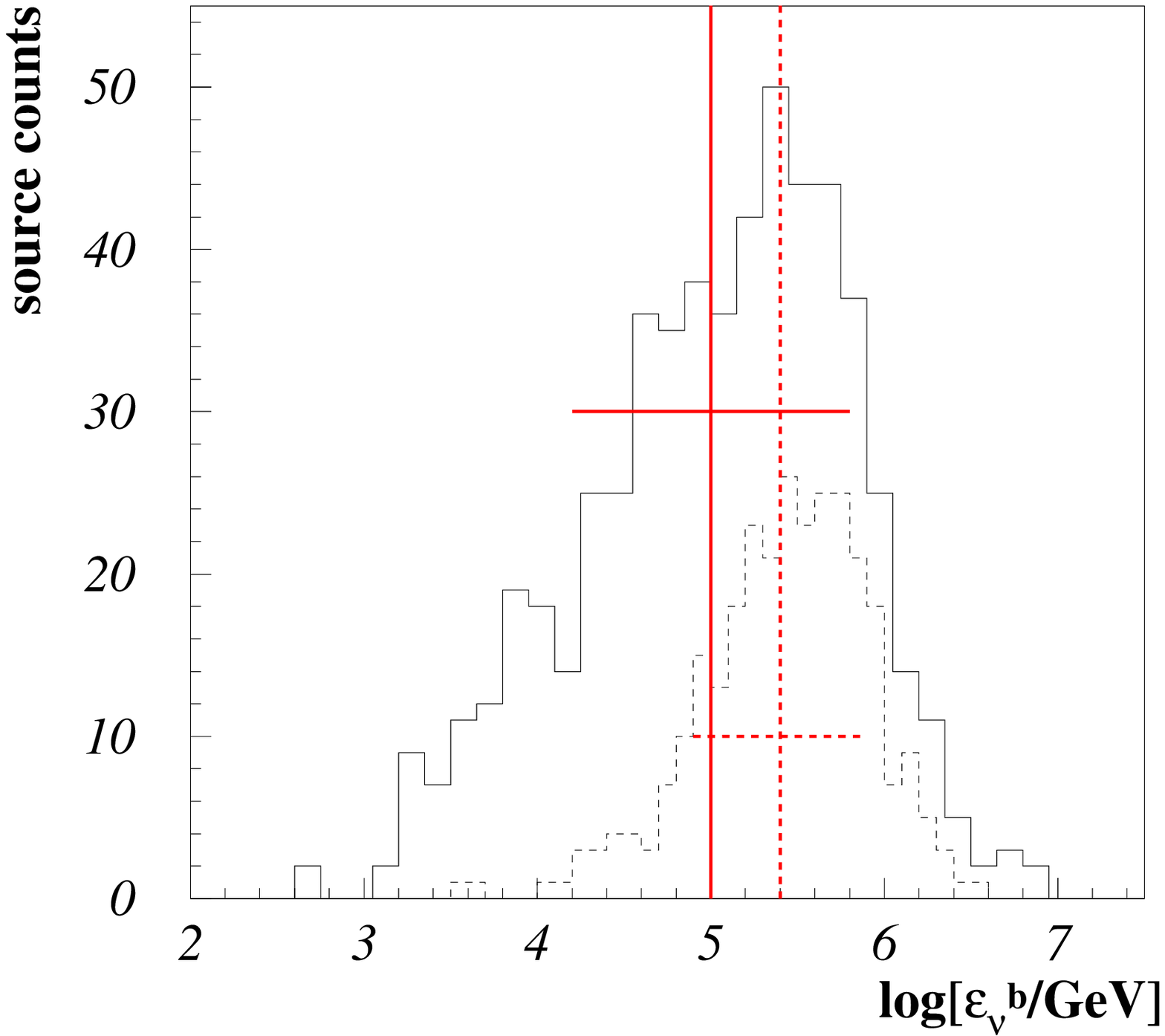}
\end{picture}\par
\caption[]{Distribution of first break energy. Solid
  line: variability sample; Dashed line: lag sample.\\[0.0cm]}
\label{enb}
\end{minipage}
\parbox{0.5cm}{\quad}
\begin{minipage}[t][9.5cm][b]{6.7cm}
\begin{picture}(6.7,3.5)
\includegraphics[width=8.cm]{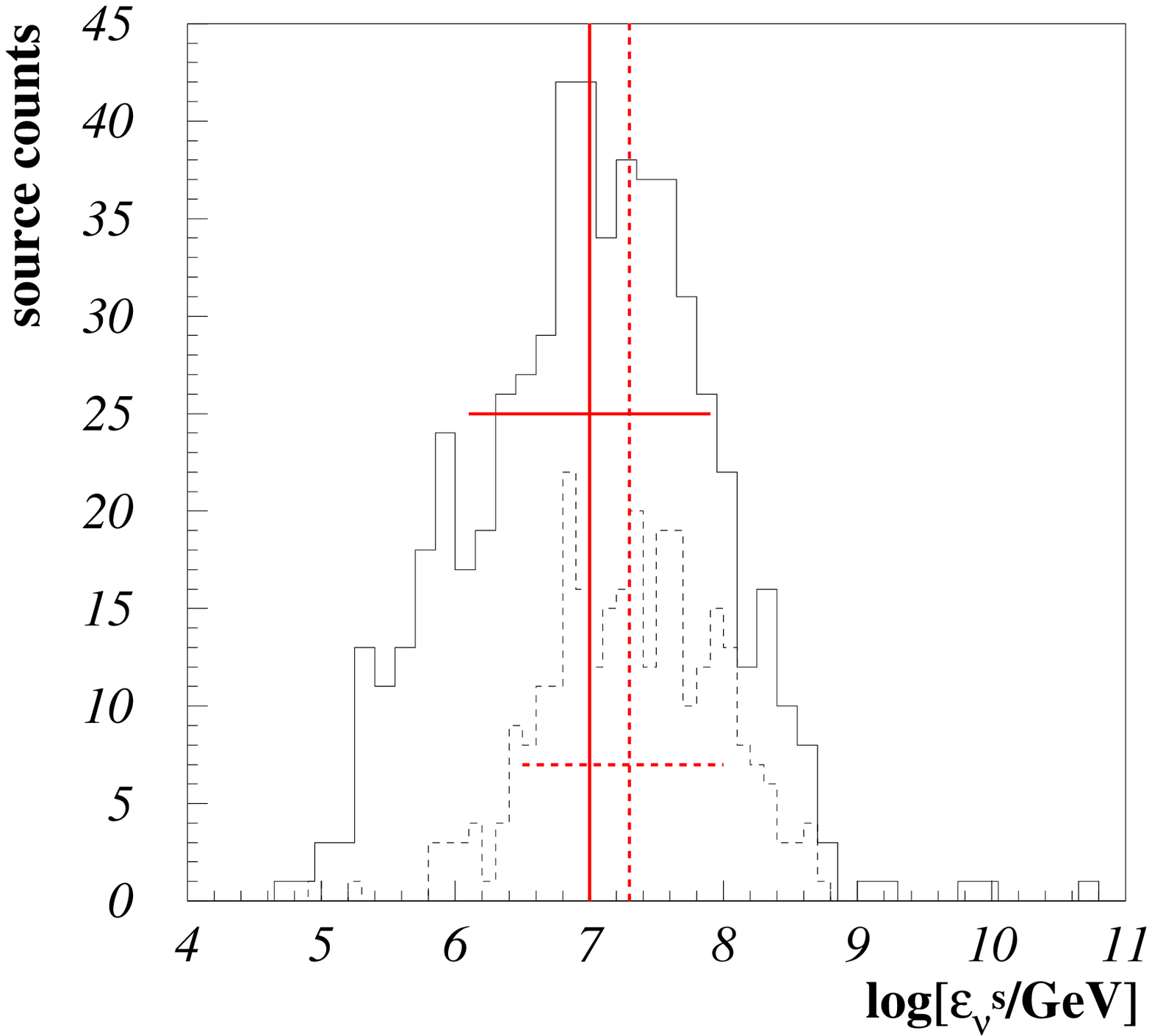}
\end{picture}\par
\caption[]{Distribution of second break energy. Solid
  line: variability sample; Dashed line: lag sample.\\[0.04cm]}
\label{ens}
\end{minipage}\hfill
\end{figure}

\begin{figure}
\setlength{\unitlength}{1cm}
\begin{minipage}[t][9.5cm][b]{6.7cm}
\begin{picture}(6.7,3.5)
\includegraphics[width=8.cm]{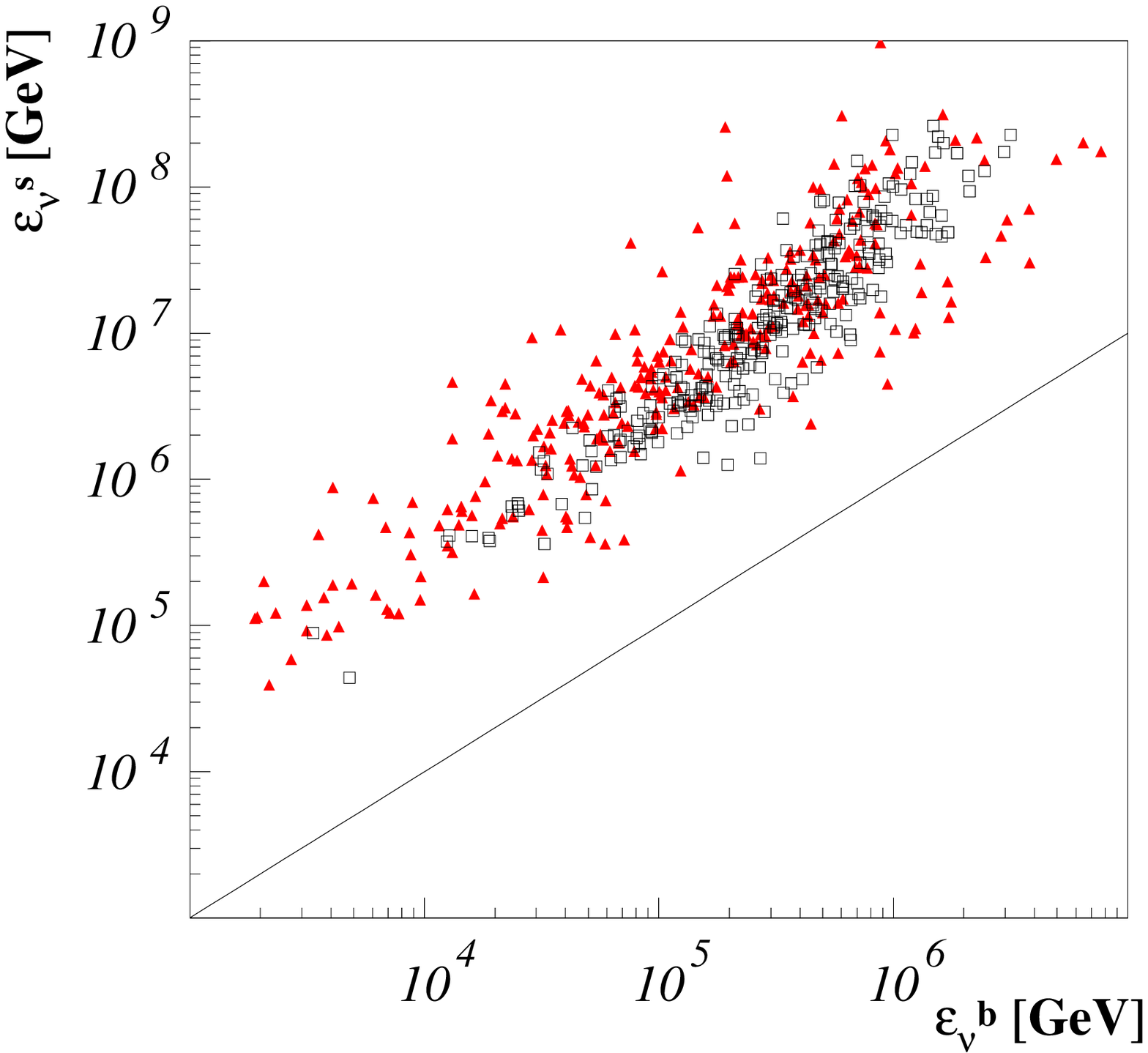}
\end{picture}\par
\caption{$\enb$ versus $\ens$. A correlation between the two break energies is
  present due to a correlation of Equ.~\ref{enb_equ} and \ref{ens_equ}. Black squares represent
the correlation for the lag sample, red triangles represent values for the
variability sample. The black line shows $\enb=\ens$.}
\label{enb_ens}
\end{minipage}
\parbox{0.5cm}{\quad}
\begin{minipage}[t][9.5cm][b]{6.7cm}
\begin{picture}(6.7,3.5)
\includegraphics[width=8.cm]{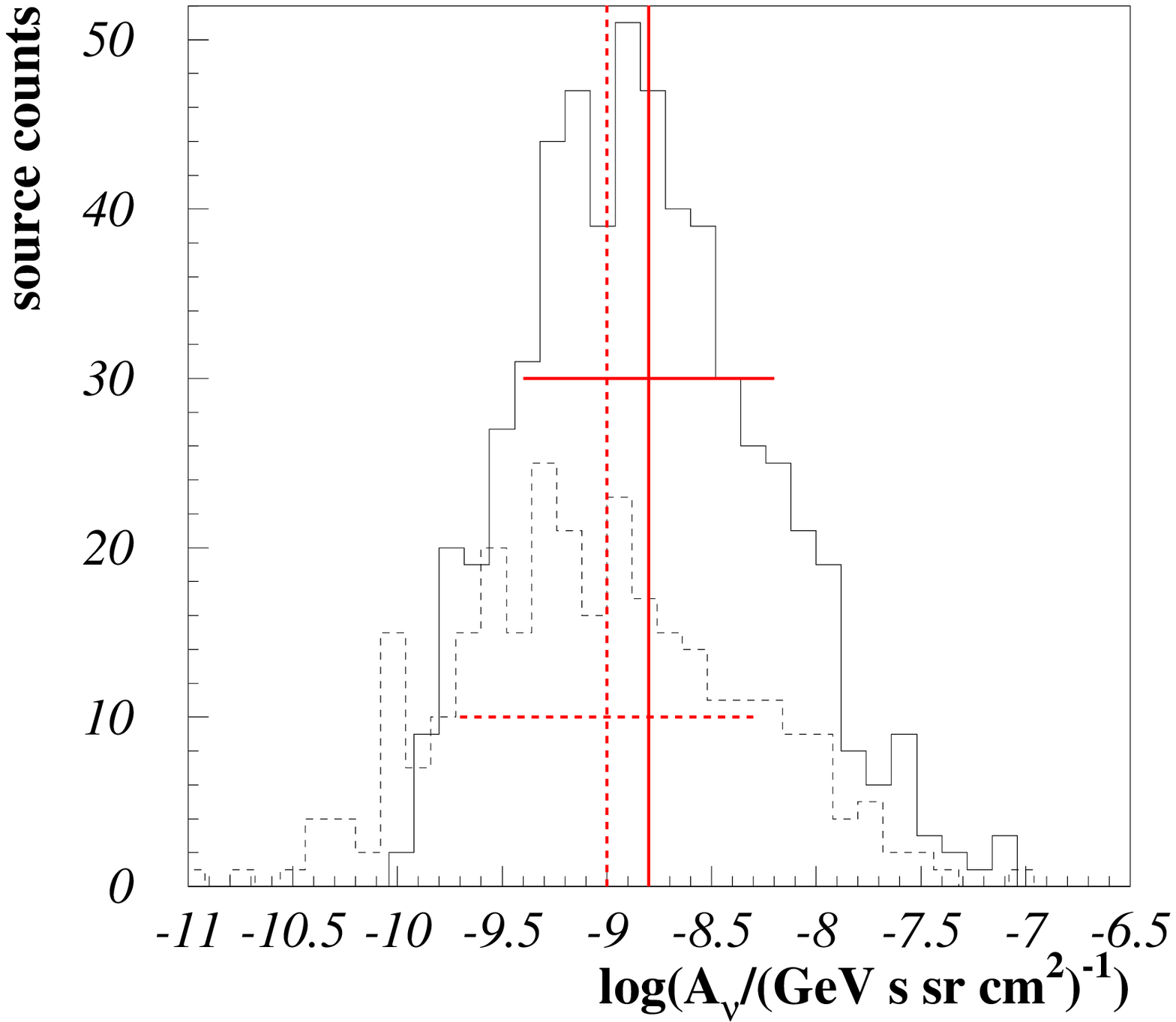}
\end{picture}\par
\caption[]{Distribution of the normalization of the single burst neutrino
  spectra, solid line: variability sample; dashed line: lag sample.\\[0.95cm]}
\label{flux}
\end{minipage}\hfill
\end{figure}

There resulting break energies are shown in Fig.~\ref{enb} and
\ref{ens}. Both samples scatter widely in both break energies, so that the
errors in the average values, ${\overline{\enb}}^{var}=5.0\pm0.8$ and
${\overline{\enb}}^{lag}=5.4\pm0.5$ as well as ${\overline{\ens}}^{var}=7.0\pm0.9$ and
${\overline{\ens}}^{lag}=7.3\pm0.7$ allow a deviation of about up to an order of magnitude
from the mean values. Scatter plots of $\enb$ and $\ens$ also show that there
is a relatively strong correlation between the first and the second break
energy in both samples with $\enb\propto \ens$, see Fig.~\ref{enb_ens}. 

Fig.~\ref{flux} shows the normalization of the variability (solid line) and
lag spectra (dashed line). The distribution of the variability
sample shows a peak with a mean value at $\log[A'_{\nu}/$GeV
    s$^{-1}$ sr$^{-1}$ cm$^{-2}]\sim -8.8\pm 0.6$, while the
maximum for the lag sample is slightly lower,  $\log[{A'}_{\nu}/$GeV
    s$^{-1}$ sr$^{-1}$ cm$^{-2}]\sim -9.0\pm0.7$. 
\clearpage

The bursts that are most interesting in the presented sample in order to seek
for an optimal spectrum for neutrino detection with a large volume neutrino
Cherenkov telescope are those bursts with low first break energies - that
means that the spectrum decreases with a relatively hard spectral index
$\alpha_{\nu}$ up to high energies, $\en>10^{5}$. Furthermore, high
normalization bursts, $A_{\nu}'>\overline{A}_{\nu}'$ will have a strong
influence on the shape of the coincidence spectrum. The coincidence spectra
are being discussed in section~\ref{coincidence}.
\subsection{Subsample from variability for AMANDA data analysis}
The last four years of BATSE operation time, 1997-2000, coincides with the
beginning years of the AMANDA experiment. That implies the possibility of an examination of
the neutrino signal from a BATSE subsample. $105$ BATSE bursts lie within
AMANDA's field of view ($\sim$ northern hemisphere) as is examined in~\cite{mike}. Since the lag sample only contains bursts
from before 1997, it cannot be used to examine any AMANDA burst. The
variability sample however contains $82$ of the $105$ bursts. The parameter distributions
of these bursts will not be discussed in more detail, but they are displayed in appendix 
\ref{amanda_plots}.
\section{Between coincidence and average predictions \label{coincidence}}

\begin{figure}
\setlength{\unitlength}{1cm}
\begin{minipage}[t][9.5cm][b]{6.7cm}
\begin{picture}(6.7,3.5)
\includegraphics[width=8.cm]{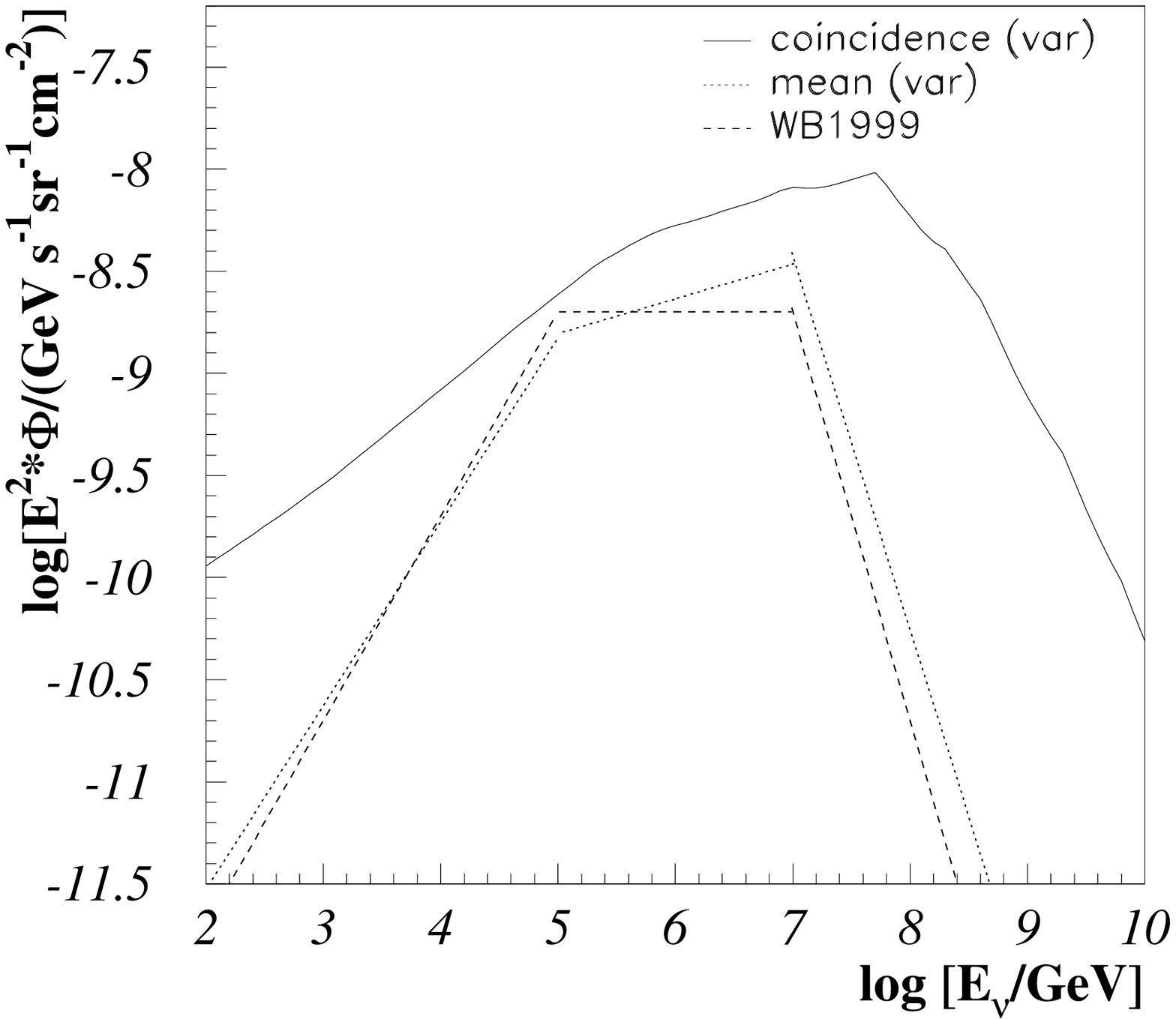}
\end{picture}\par
\caption{Comparison of coincidence and average spectrum (variability).}
\label{var_sum}
\end{minipage}
\parbox{0.5cm}{\quad}
\begin{minipage}[t][9.5cm][b]{6.7cm}
\begin{picture}(6.7,3.5)
\includegraphics[width=8.cm]{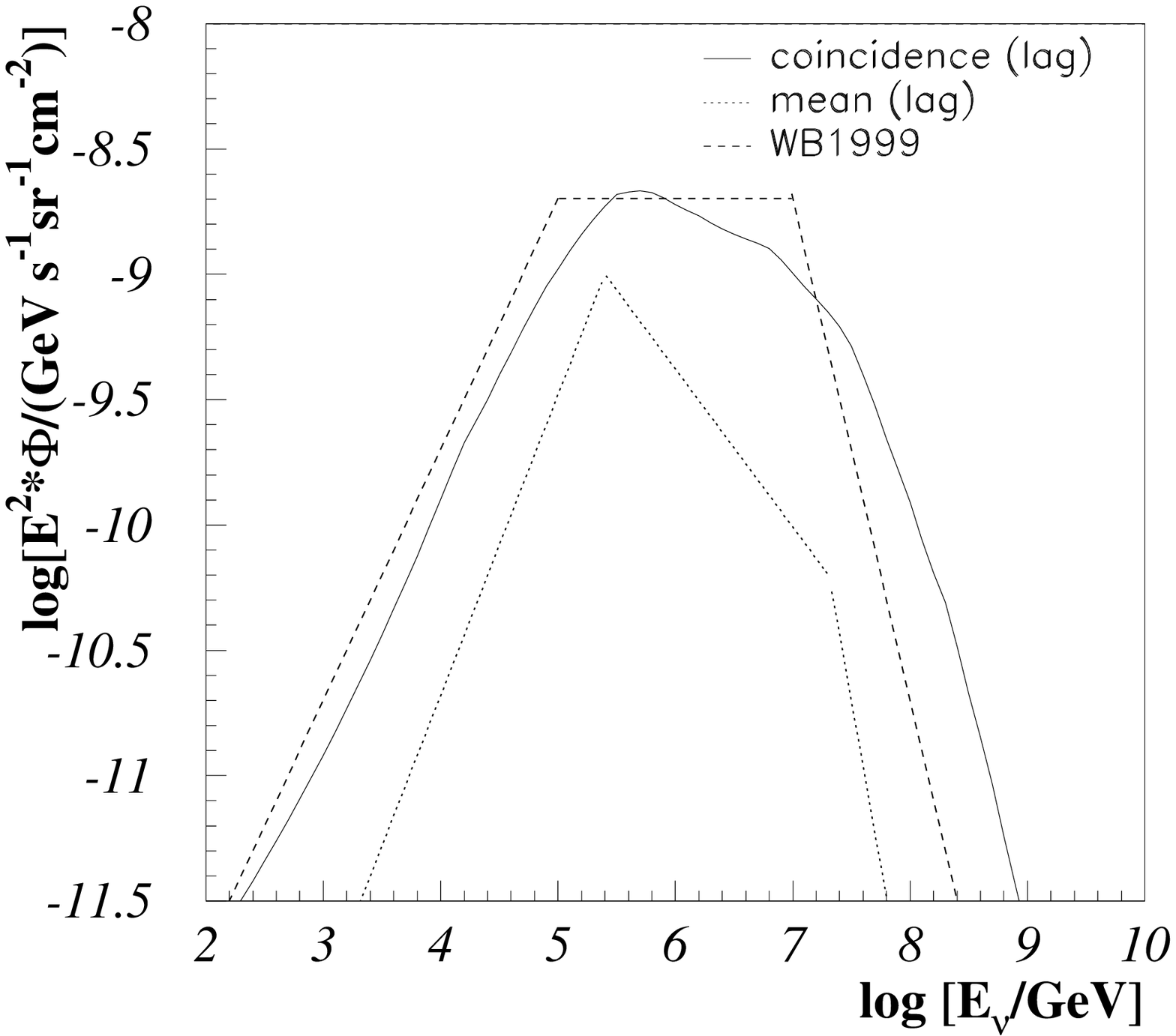}
\end{picture}\par
\caption{Comparison of coincidence and average spectrum (lag).}
\label{lag_sum}
\end{minipage}\hfill
\end{figure}

\begin{figure}
\centering{
\includegraphics[width=12cm]{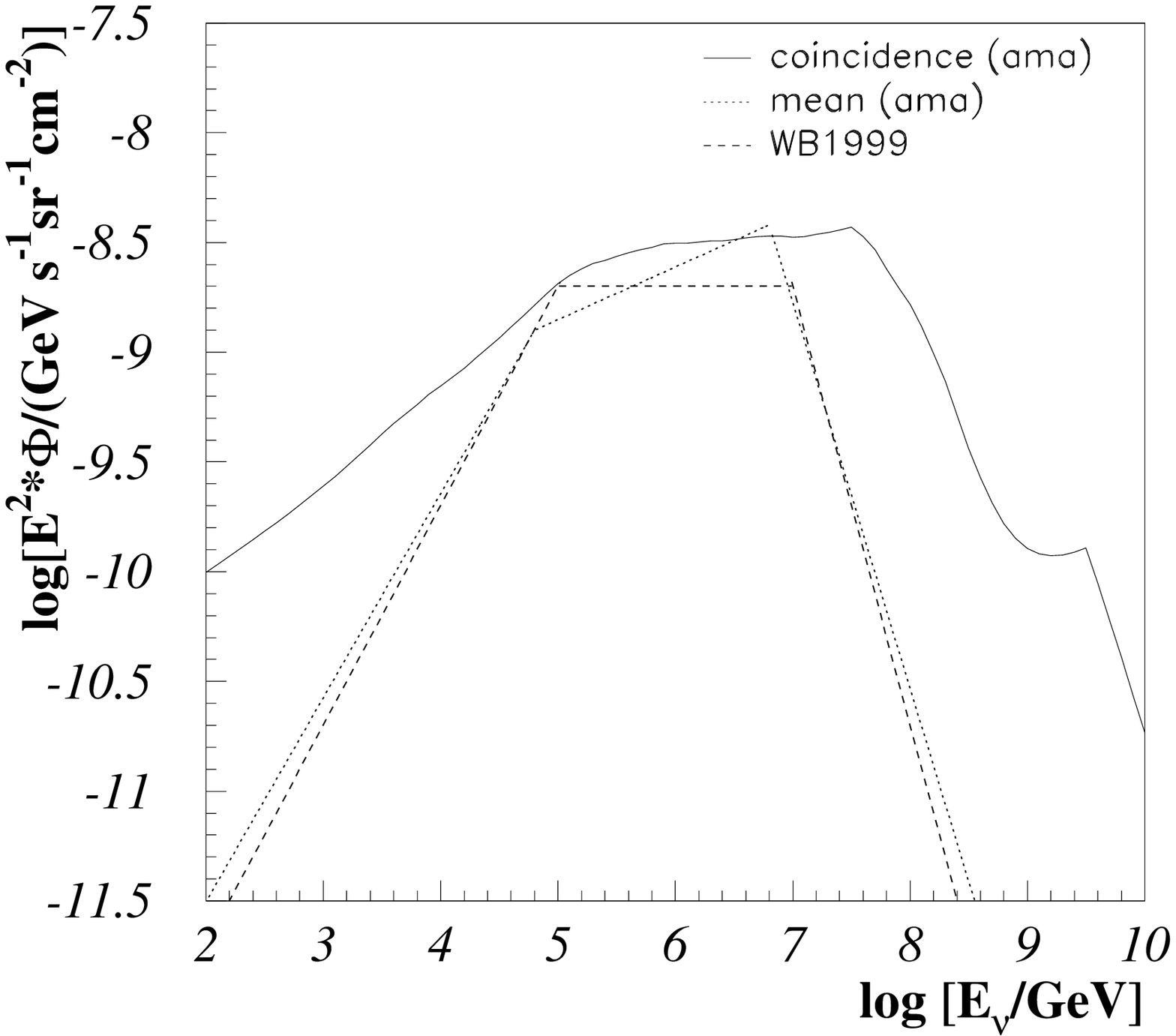}
\caption{Comparison of coincidence and average spectrum (AMANDA-variability subsample).}
\label{ama_sum}
}
\end{figure}

\begin{figure}
\setlength{\unitlength}{1cm}
\begin{minipage}[t][11.2cm][b]{6.7cm}
\begin{picture}(9.7,8.5)
\includegraphics[width=8.cm]{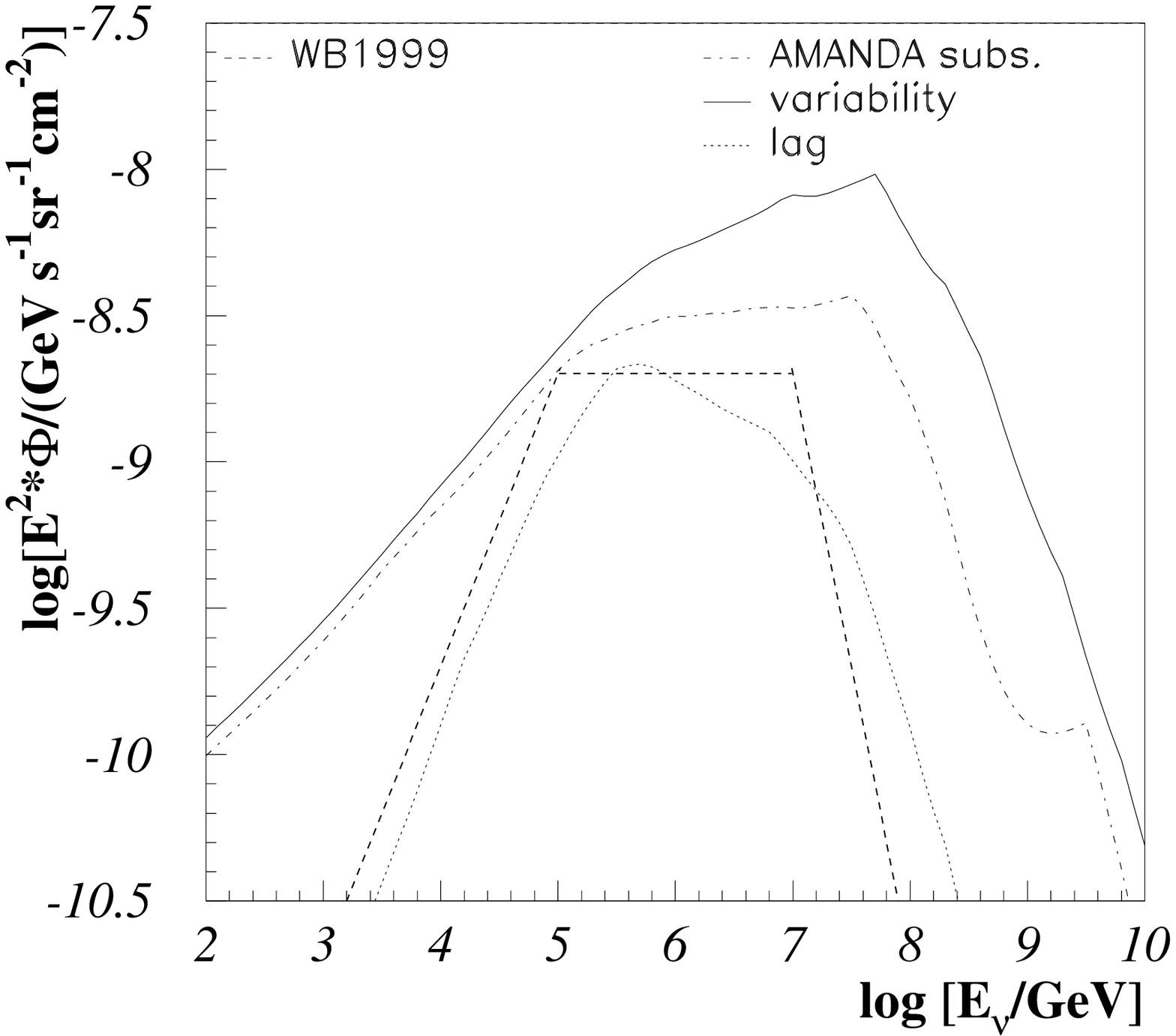}
\end{picture}\par
\caption[]{$292$ bursts (lag, dotted line), $568$ bursts from variability method
  (solid line) and a subsample of $82$ bursts from the variability sample which can be analyzed with
  AMANDA (dot-dashed line). The variability and lag sample coincidence spectra differ clearly from
  each other and also from the average, diffuse Waxman/Bahcall
  prediction. This is a strong indication that it is necessary to treat bursts
  individually in an analysis in order to optimize the signal expectation.\\[-1.4cm]}
\label{static_sum}
\end{minipage}
\parbox{0.5cm}{\quad}
\begin{minipage}[t][11.2cm][b]{6.7cm}
\begin{picture}(9.7,8.5)
\includegraphics[width=8.cm]{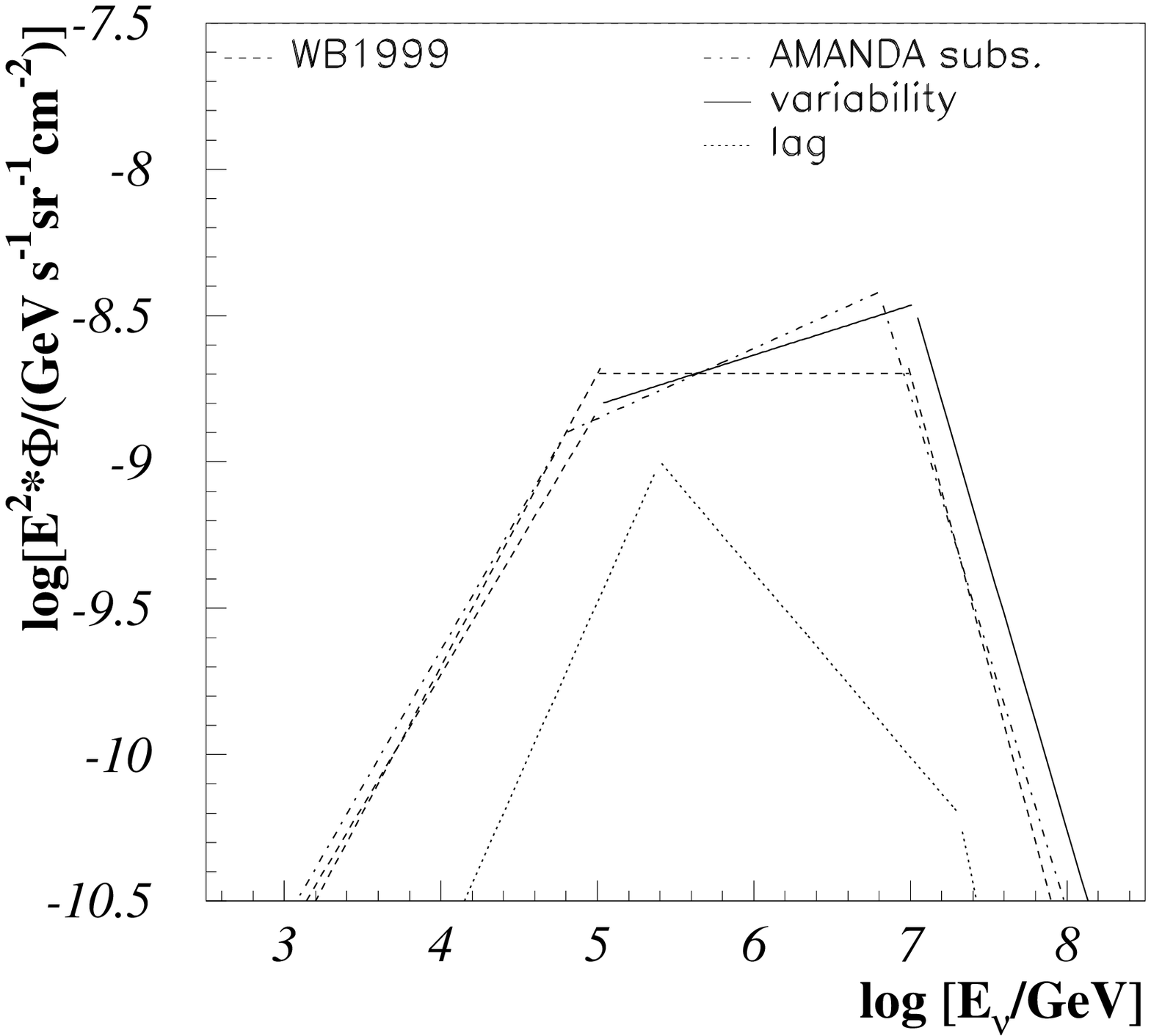}
\end{picture}\par
\caption[]{Summary of the mean spectra of all samples. Major differences can be
  seen between the variability (solid line) and lag sample (dotted line),
  mostly due to differences in the second spectral index and
  normalization. All calculated mean spectra differ significantly from the
  average diffuse prediction made by Waxman and Bahcall (dashed line). The
  AMANDA subsample is represented by the dot-dashed line.}
\label{mean_spectrum}
\end{minipage}\hfill~\\[0.5cm]
\end{figure}

For each burst $i$ in a sample, a prediction of the prompt neutrino flux from
this source $\Phi_i$ can be made as described above. To analyze GRBs with a
large volume neutrino telescope like AMANDA, a sample of bursts is analyzed to
increase the signal rate  which makes it interesting to look at a
coincident spectrum of burst samples. The total flux $\Phi$ is presented in the
following form in this calculation,
\begin{equation}
\Phi=\frac{\sum_{i}^{n}\Phi_i}{n}\,.
\end{equation}
Here, $n$ is the total number of bursts in the sample.

In this section, the coincidence spectra of each examined sample will be
described in more detail. Fig.~\ref{var_sum} shows the expected energy spectrum for
the variability sample, the solid line representing the coincidence spectrum,
the dotted line displaying the average spectrum calculated from the mean
parameters as given in table \ref{average_params}. There are large differences
between the two spectra, especially concerning the spectral indices at low and
high energies. The differences between coincidence and average spectrum in the
lag sample are rather in the normalization than in the shape of the
spectra, see Fig.~\ref{lag_sum}. The reason for these deviations lies in the contribution
of individual bursts parameters far from the average. These bursts are not
considered in average calculations at all, but can be responsible for a
significant change in the spectral shape and normalization. The average and
coincidence flux for the AMANDA subsample is shown in Fig.~\ref{ama_sum}. Both versions of the energy spectra follow the larger sample
with smaller deviations, because the distributions of the subsample show a
similar behavior as the whole sample (see appendix). What can also be seen is
that for all samples both average and coincidence spectrum differ from the
spectrum that is considered as the standard diffuse spectrum. The differences
between diffuse and lag spectrum are quite small, but the spectrum of the
variability sample has significantly different features compared to the
diffuse spectrum.

To compare the different samples to each other, the coincidence fluxes of all
samples are shown in Fig.~\ref{static_sum}. The solid line shows the
variability sample spectrum and the AMANDA subsample is represented by the
dot-dashed line. The lag sample is given by the dashed line. The diffuse spectrum from
\cite{WB} is indicated by the dashed line as a comparison. It can be seen at
first glance that the two total samples show a very different behavior. Here, the difference from the
average diffuse spectrum is of special interest, since this shows that
coincidence spectra do not seem to be treatable as average diffuse spectra as
easily. It can be seen that apart from the global normalization of the
spectrum, the slope of especially the second spectral index is relevant for
the overall normalization of the spectrum.  A similar effect can be seen when
comparing the average spectra of the samples to each other and to the diffuse
spectrum, see Fig.~\ref{mean_spectrum}. Here, however, the average
variability spectrum agrees relatively well with the diffuse spectrum while
there is quite a large difference between lag and diffuse prediction.
\section{Estimate of Cherenkov detection rate}
An estimate of the detection rate $R$ can be given by folding the expected
neutrino flux at Earth as it has been calculated above with the probability of
the detection of the neutrino,
\begin{equation}
R(E_{\min},\theta)=\int_{E_{\min}}P_{\nu\rightarrow l}(E_{\nu},E_{\min})P_{shadow}(\theta,E_{\nu}) \Phi_{\nu}dE_{\nu}\,.
\end{equation}
Here, $P_{\nu\rightarrow l}(E_{\nu},E_{\min})$ is the probability that a
neutrino interacts with a nucleus to produce either a muon - $l=\mu$ - or an electromagnetic
cascade - $l=$cascade - which is detectable in a large volume neutrino
detector. It can be written as
\begin{equation}
P_{\nu\rightarrow l}=N_A\,\int_{E_{\min}}^{E_{\nu}}dE_{l}\,\frac{d\sigma}{dE_{l}}r_{l}(E_l,E_{\min})
\end{equation}
where $N_A$ is Avogado's constant, $r_l$ is the range of the produced muon -
$l=\mu$ - or cascade - $l=$cascade - within detection range and $d\sigma/dE_{l}$ is
the differential charged current cross section for $N\,\nu$
interactions. $E_{\min}$ is the energy threshold of the detector for an event
detection. The cross section is determined using the parton distribution
functions given by \cite{pdflib}, where the model of \cite{grv} is used.

$P_{shadow}$ is the probability that the neutrino is absorbed by the
Earth. It is given as
\begin{equation}
P_{shadow}=\exp\left(-X(\theta)/\lambda\right)\,.
\end{equation}
The neutrino absorption length $X(\theta)$ is dependent on the angle of the
incoming neutrino towards the nadir, $\theta$. It is determined by the
distance that the neutrino travels through Earth and the Earth's density using
~\cite{earthmodel} and the atmosphere's density of the US standard atmosphere model. For a description of $X(\theta)$ and $\lambda$ see for example \cite{gaisser}.

\begin{figure}[ht]
\centering{
\includegraphics[width=12cm]{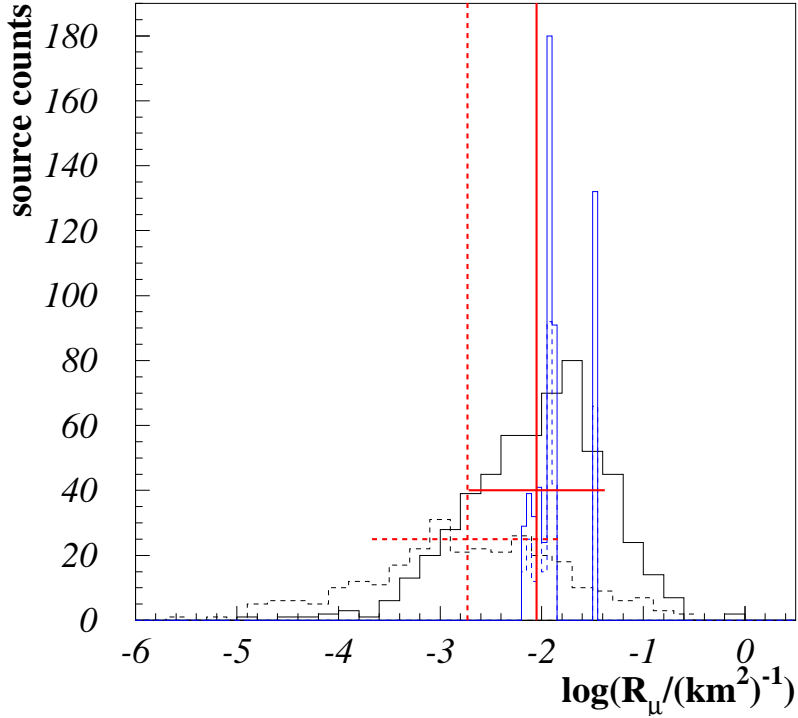}
\caption{Muon neutrino rates for the single bursts using a threshold energy of $E_{\min}=100$~GeV. The
  rates are given per burst and per square kilometer. On average, the neutrino
  flux rates determined in the lag sample are lower than the ones for the
  variability sample ($\log(R_{\mu}^{lag}/\mbox{km}^{-2})= -2.73\pm 0.94$ compared to
  $\log(R_{\mu}^{var}/\mbox{km}^{-2})= -2.05\pm 0.67$). The blue lines represent the same burst
  samples (solid: variability; dashed: lag) with the Waxman-Bahcall spectrum
  as input flux. All parameters except the angle are fixed as described above.}
\label{rates}
}
\end{figure}

\begin{figure}[ht]
\centering{
\includegraphics[width=12cm]{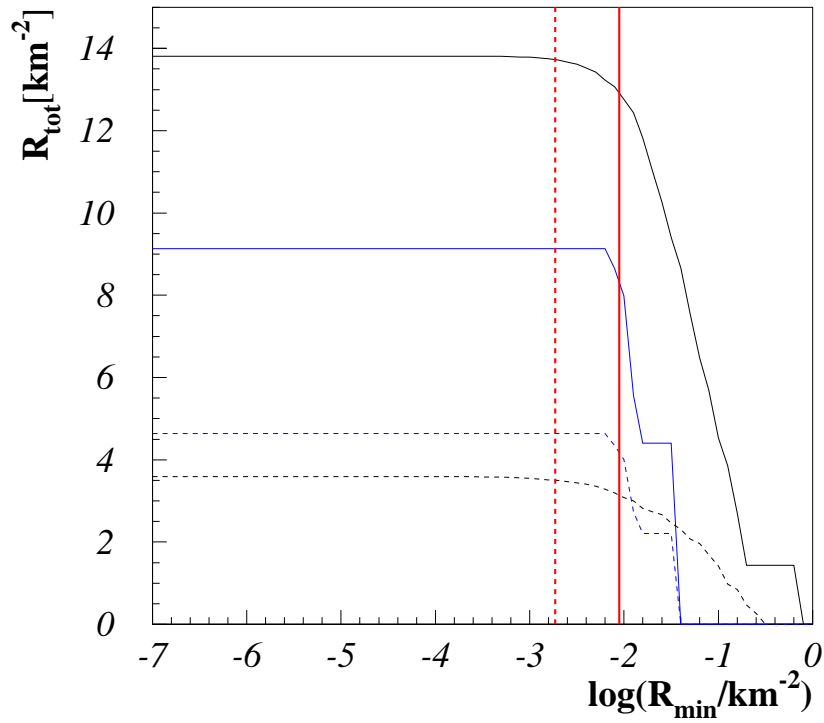}
\caption{Total muon rate of the complete sample (variability: solid black
  line, lag: dashed black line). The lower summation limit is plotted on the
  x-axis. Only a few sources below average (red lines) contribute to
  the rate. As a comparison, the samples have been regarded using the
  Waxman-Bahcall spectrum as an input flux (blue lines), solid: variability;
  dashed: lag.}
\label{total_rate}
}
\end{figure}

The rate of neutrino induced muons per burst is displayed in Fig.~\ref{rates}, the solid line representing bursts from the variability
sample, the dashed line showing the bursts from the lag sample. It can be seen
that there is no burst yielding the probability of a whole event in a square kilometer detection
array. This indicates that a source stacking method is useful for an analysis
to get a higher significance for an actual detection. Furthermore, the fits
from the variability sample yields more neutrinos on average than the ones
from the lag method. As it has already been stated in \cite{guetta}, it can be
confirmed in this analysis that there are a few bursts with rates above
average which will dominate the flux from the sample. Bursts below average
give a small contribution to the total signal. This is demonstrated in Fig.~\ref{total_rate}: The y-axis shows the total muon neutrino rate per sample,
\begin{equation}
R_{tot}(R_{i_{\min}})=\sum_{i_{\min}}^{n} R_{i}\,.
\end{equation}
Here, $i_{\min}$ is the lower summation index and $R_{i}$ are the
corresponding single burst rates. It is successively shifted to
higher indices, $i_{\min}=1,\,2,\,...,\,n$ with $n$ as the number of sources in
the sample. The sources are arrange such as the lower rates have lower
indices. For instance, the lowest rate in a sample gets the index $i=1$, $R_{i}$,
etc. Thus, by increasing $i_{\min}$, less and less sources are
considered, starting by removing the least luminous ones. Fig.~\ref{total_rate} shows $R_{tot}$ versus
$R_{i_{\min}}$. It can be seen that bursts with very low rates do not
contribute significantly. Approaching the mean value of the distributions,
however, the total rate starts to decrease rapidly. This behavior is seen in
both samples. If the Waxman-Bahcall (WB) flux is used as the input spectrum
(blue lines), the
flux is constant at first as well, but plunges down extremely, because the
rate distribution does not scatter very much as is seen in Fig.~\ref{rates}. The variability sample for instance shows a drop-off about an
order of magnitude before the WB-flux would drop of using the same
sources. At this point, the flux has already decreased by one event where all
events are still captured in the WB-scenario. This one event comes from the
sources below average in the variability sample - the remaining 12.7 events
result from the upper part of the spectrum. This shows that most of the
contribution actually comes from bursts above average. Compared to the input
of real parameters, the WB-scenario would give about 4.5 events less. The lag
sample in contrast would only yield 3.6 events compared to 4.7 events
in the WB-scenario. Mean and total values for the different scenarios are
given in table~\ref{average_rate_params}.

\begin{table}[ht]
\centering{
\begin{tabular}{|c||c|c|}\hline
&mean $\nu_{\mu}$ rate [km$^{-2}$]&log total $\nu_{\mu}$rate [km$^{-2}$]\\
    \hline\hline
variability&$10^{-2.05\pm0.67}$&13.8\\\hline
WB [var. sample]&$10^{-1.86\pm0.23}$&9.12\\ \hline
lag&$10^{-2.73\pm0.94}$&3.6\\\hline
WB [lag sample]&$10^{-1.86\pm0.23}$&4.7\\ \hline
AMANDA subsample&$10^{-2.05 \pm 0.59}$&1.54\\ \hline
\end{tabular}
\caption{Mean neutrino spectra parameters for the three samples, variability
  (568 bursts, single burst parameters and WB), lag (292, single burst parameters and WB) and the variability subsample of AMANDA bursts (82 bursts). The
  standard deviation to the mean values has been calculated as an error
  estimate.}
\label{average_rate_params}
}
\end{table}
\clearpage
\section{Conclusions and Outlook}
It is known since the detailed analysis of BATSE energy spectra that
the parameters of the Band fits are not universal, but can differ
significantly with the individual bursts. This implies that also the neutrino
spectra of single sources have very different appearances. 

The challenging task in the future will be to get a more accurate description of
the redshift estimators, since until today, the relations are still derived
from a very small sample of bursts with measured redshifts. The first redshift
measurements of Swift give hope to a significant enlargement of the sample of measured
redshifts in the near future which makes an improvement of the redshift
estimators possible. Since it is known that GRBs follow Star Formation Rate
(SFR), a strong redshift evolution is expected. Therefore, the redshifts are
an essential parameter in neutrino flux calculations and the currently
available redshift estimators are the most accurate way of dealing with
redshifts so far. Other parameters like the spectral indices can be derived
directly from the burst fits and therefore only include the errors due to
uncertainties in the measurement of the photon spectrum.

In this paper, a prediction of the neutrino flux from individual GRBs could
be made using the model of prompt neutrino emission that has been developed by
\cite{WB}. We follow the ansatz of \cite{guetta} by using individual
parameters for each burst instead of using average parameters as it has been
done by \cite{WB}. It could be shown that it is quite important to look
at GRB neutrino spectra individually to get an accurate description of a
signal from these sources. This is of special interest for large volume
neutrino telescopes as AMANDA, since an accurate description of the potential
signal is necessary in order to optimize an analysis. 
This work shows that the use of an average spectrum for analysis purposes implies the danger of a misinterpretation
of the results.
Particularly, it could
be shown that even the average spectra of a sample of bursts is likely to
differ significantly from the diffuse prediction made by \cite{WB}.

\section*{Acknowledgments}
The authors would like to thank Dan Hooper, David Band and Eli Waxman for brisk and helpful
discussions.\\
This work has been supported by the \emph{Deutsche Forschungsgemeinschaft}
(DFG).
\clearpage
\appendix
\section{Parameter distribution of the AMANDA subsample (82 bursts)
  \label{amanda_plots}}
\begin{figure}[h]
\setlength{\unitlength}{1cm}
\begin{minipage}[t][9.5cm][b]{6.7cm}
\begin{picture}(6.7,3.5)
\includegraphics[width=8.cm]{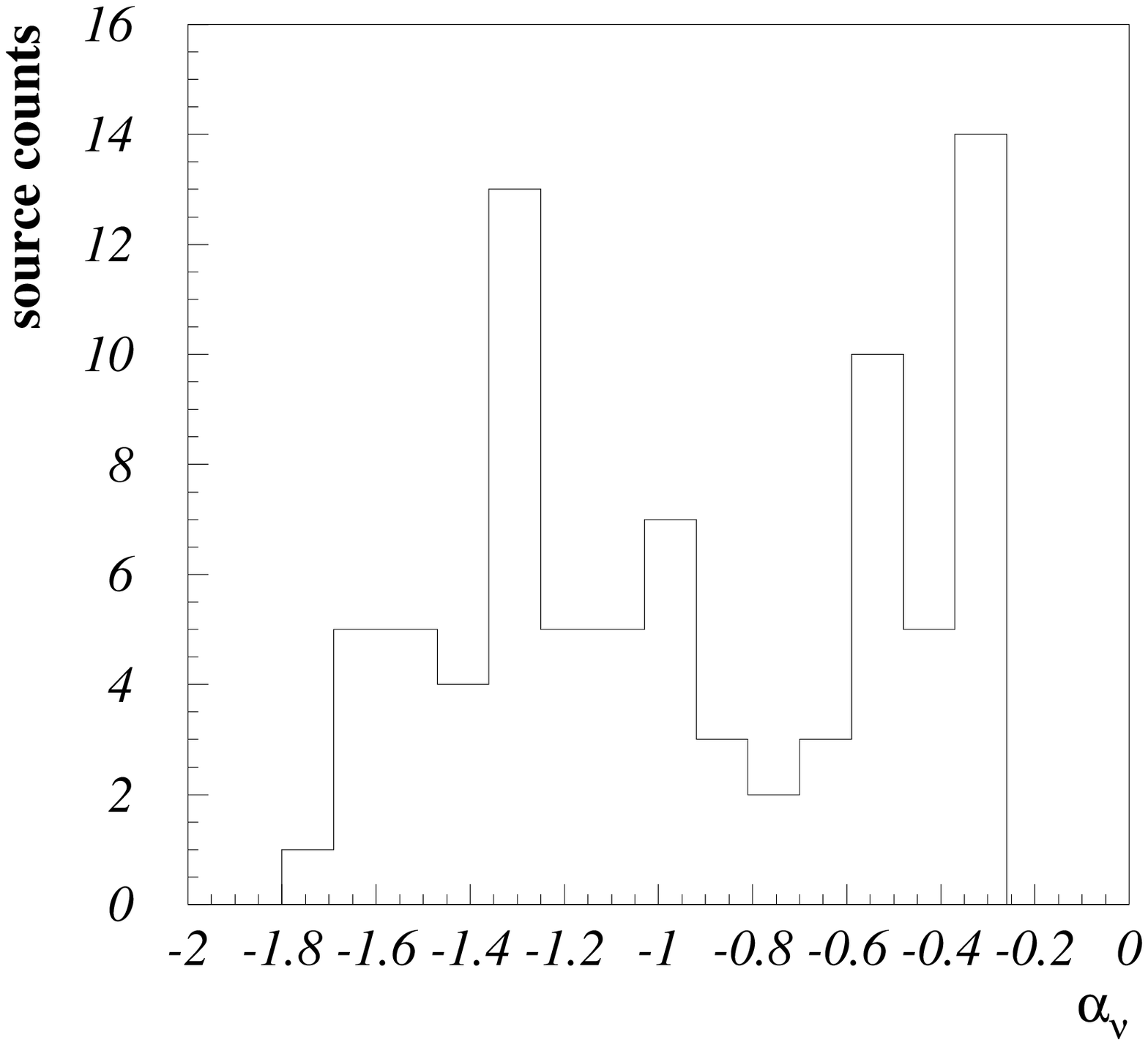}
\end{picture}\par
\caption{$\alpha_{\nu}$: first spectral index of the neutrino spectrum.}
\label{alpha_ama}
\end{minipage}
\parbox{0.5cm}{\quad}
\begin{minipage}[t][9.5cm][b]{6.7cm}
\begin{picture}(6.7,3.5)
\includegraphics[width=8.cm]{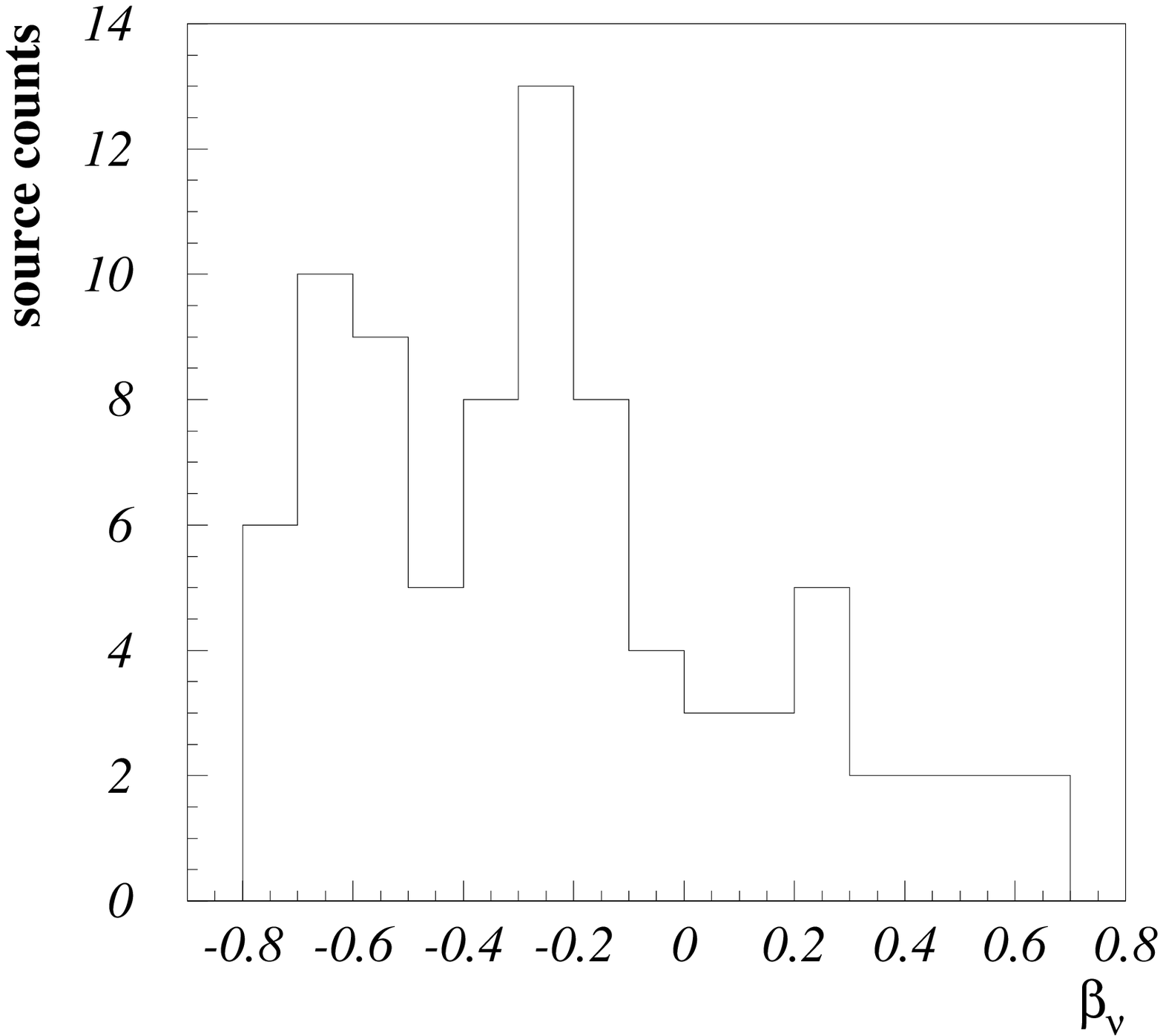}
\end{picture}\par
\caption{$\beta_{\nu}$: second spectral index of the neutrino spectrum.}
\label{beta_ama}
\end{minipage}\hfill
\end{figure}

\begin{figure}[h]
\setlength{\unitlength}{1cm}
\begin{minipage}[t][9.5cm][b]{6.7cm}
\begin{picture}(6.7,3.5)
\includegraphics[width=8.cm]{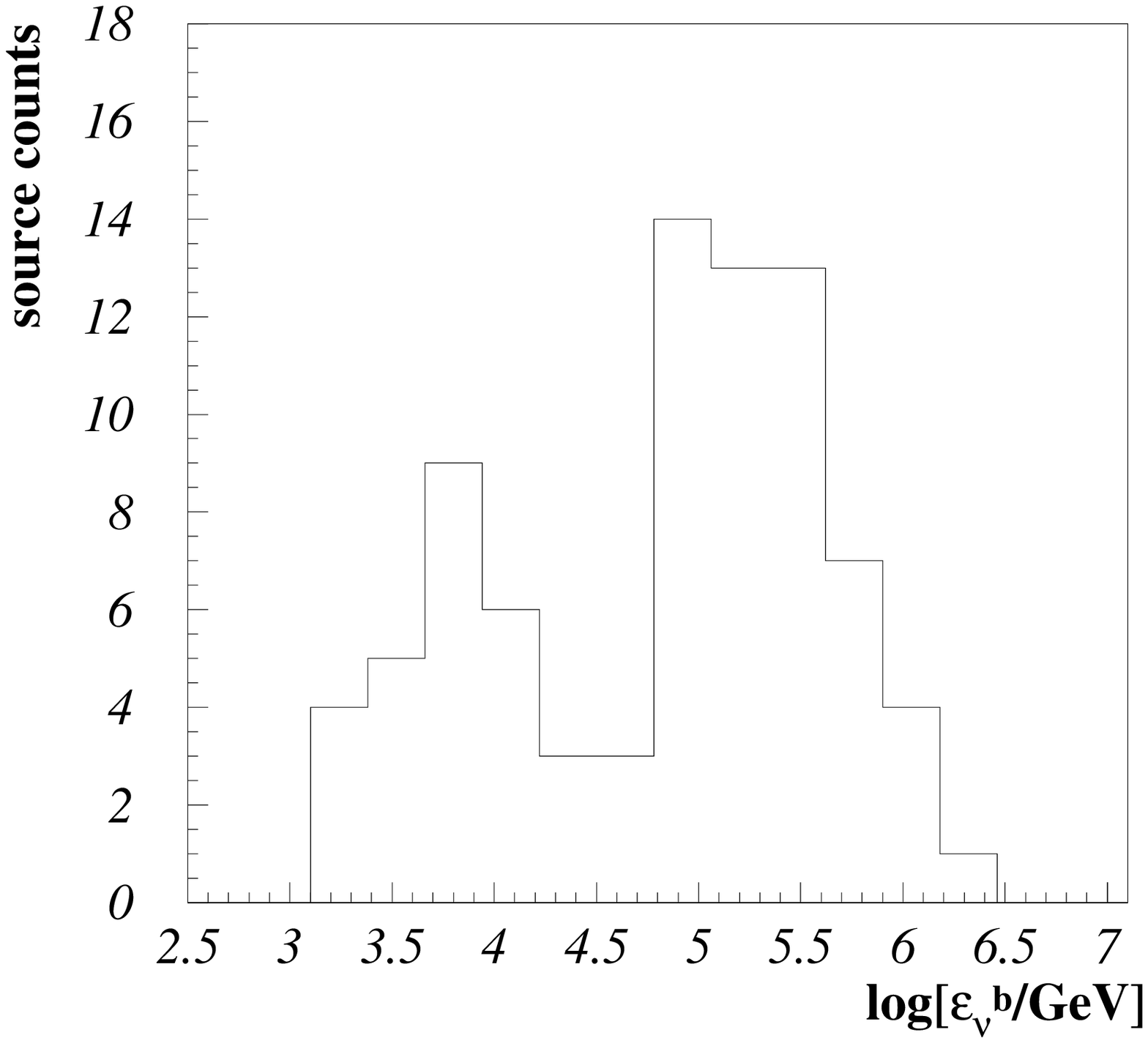}
\end{picture}\par
\caption{First break energy.}
\label{enb_ama}
\end{minipage}
\parbox{0.5cm}{\quad}
\begin{minipage}[t][9.5cm][b]{6.7cm}
\begin{picture}(6.7,3.5)
\includegraphics[width=8.cm]{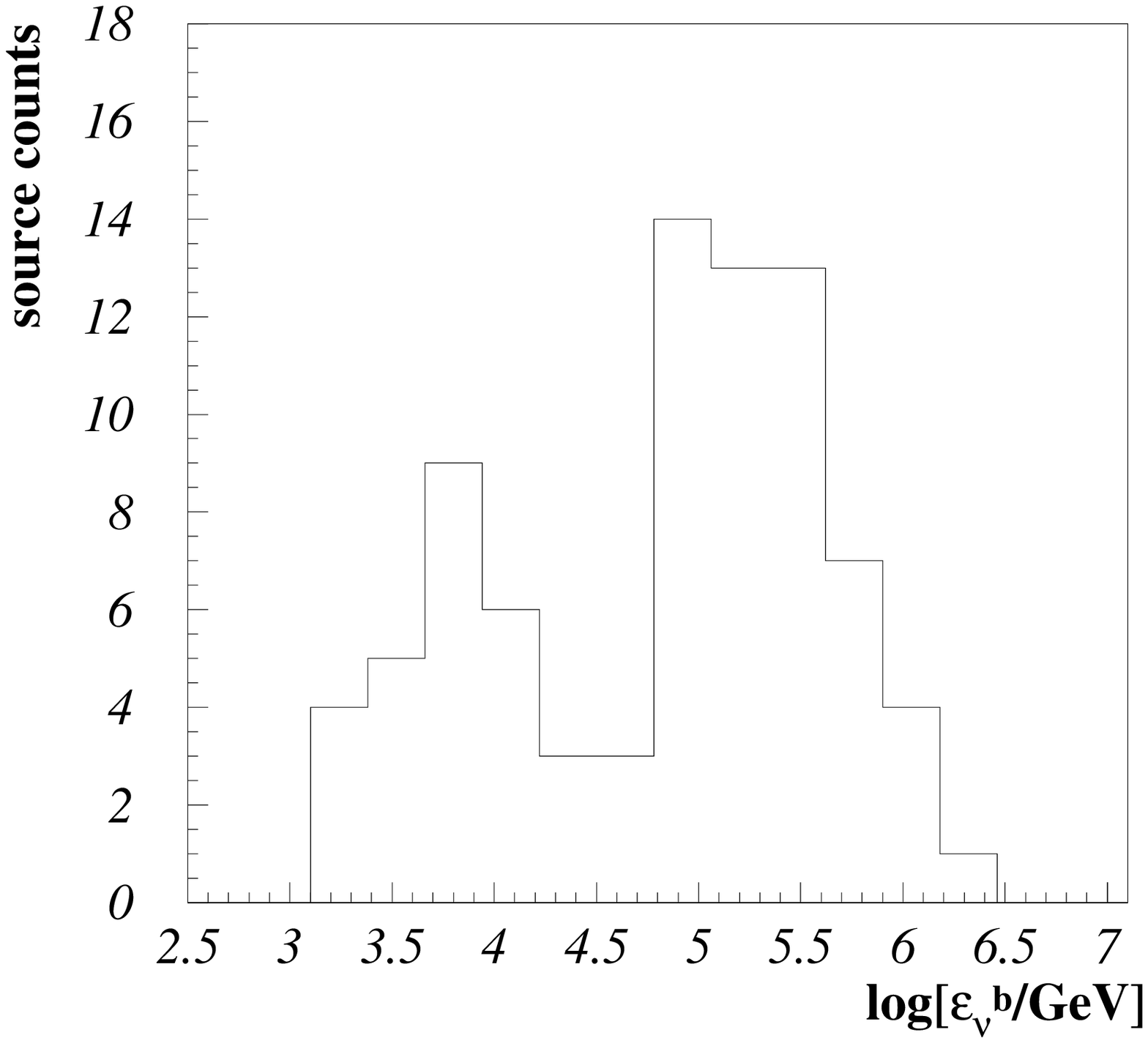}
\end{picture}\par
\caption{Second break energy.}
\label{ens_ama}
\end{minipage}\hfill
\end{figure}

\begin{figure}[h]
\setlength{\unitlength}{1cm}
\begin{minipage}[t][9.5cm][b]{6.7cm}
\begin{picture}(6.7,3.5)
\includegraphics[width=8.cm]{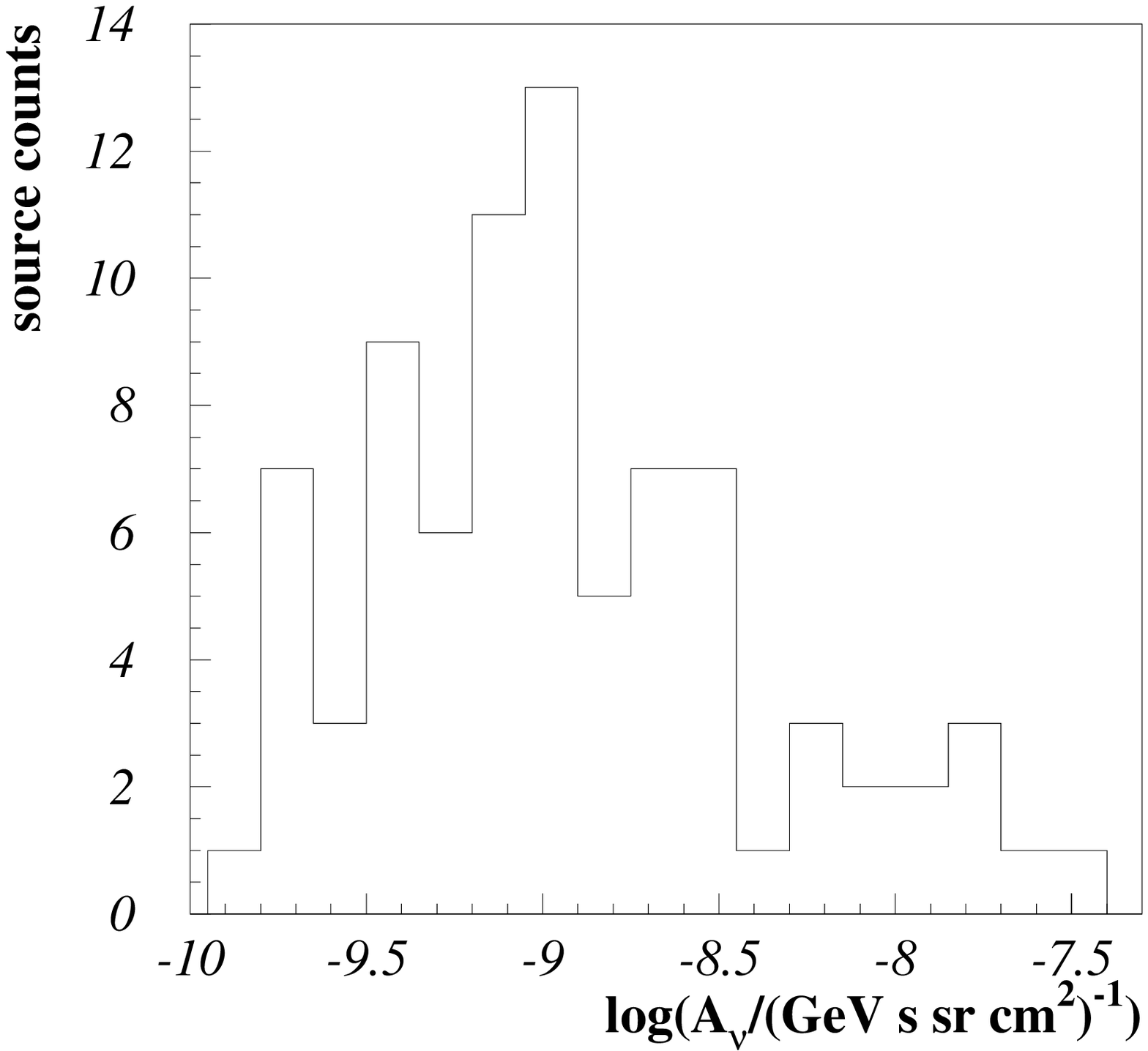}
\end{picture}\par
\caption{Normalization of the spectrum.}
\label{flux_ama}
\end{minipage}
\parbox{0.5cm}{\quad}
\begin{minipage}[t][9.5cm][b]{6.7cm}
\begin{picture}(6.7,3.5)
\includegraphics[width=8.cm]{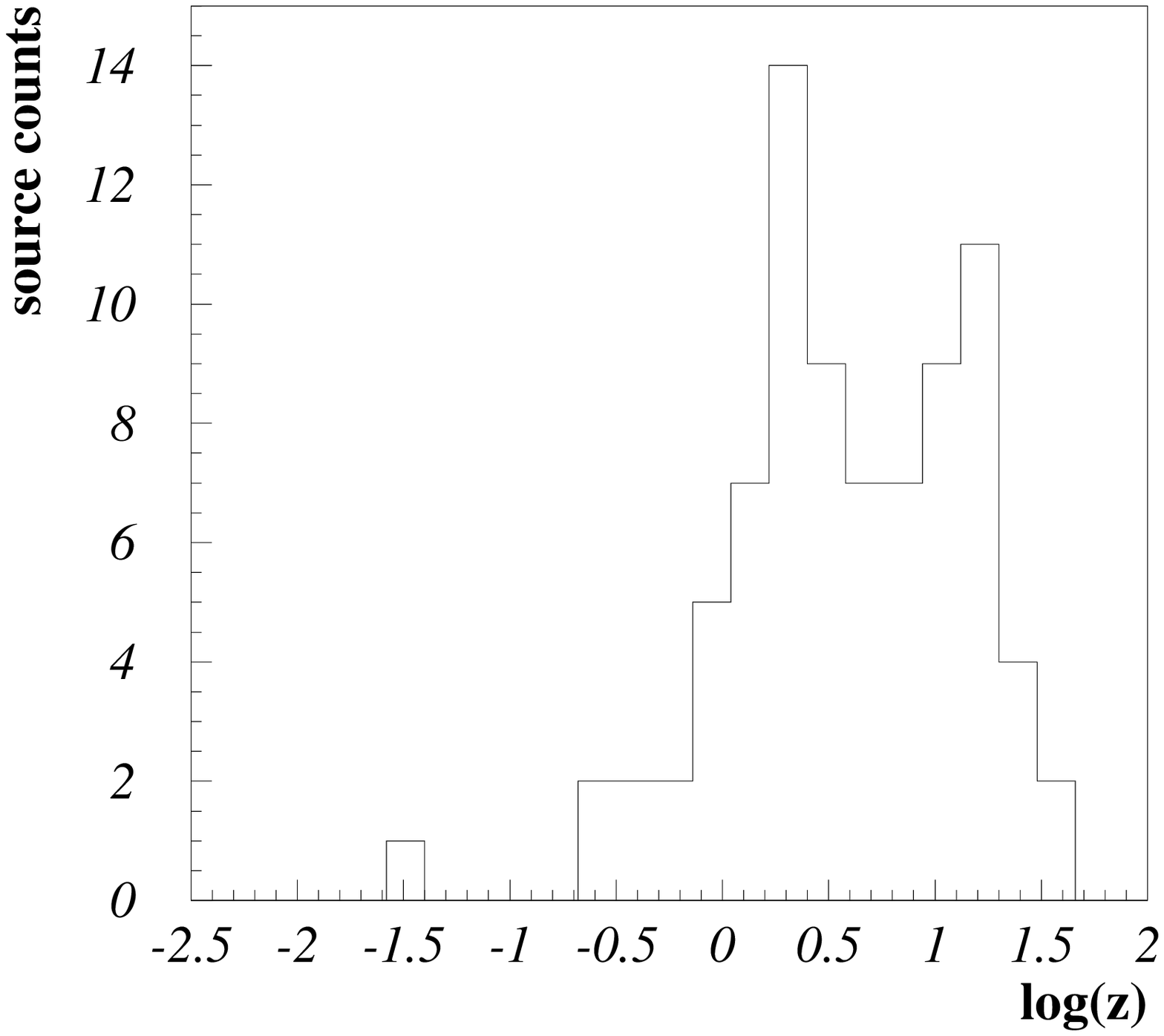}
\end{picture}\par
\caption{Redshift distribution.}
\label{logz_ama}
\end{minipage}\hfill
\end{figure}
\begin{figure}[h]
\centering{
\includegraphics[width=12.cm]{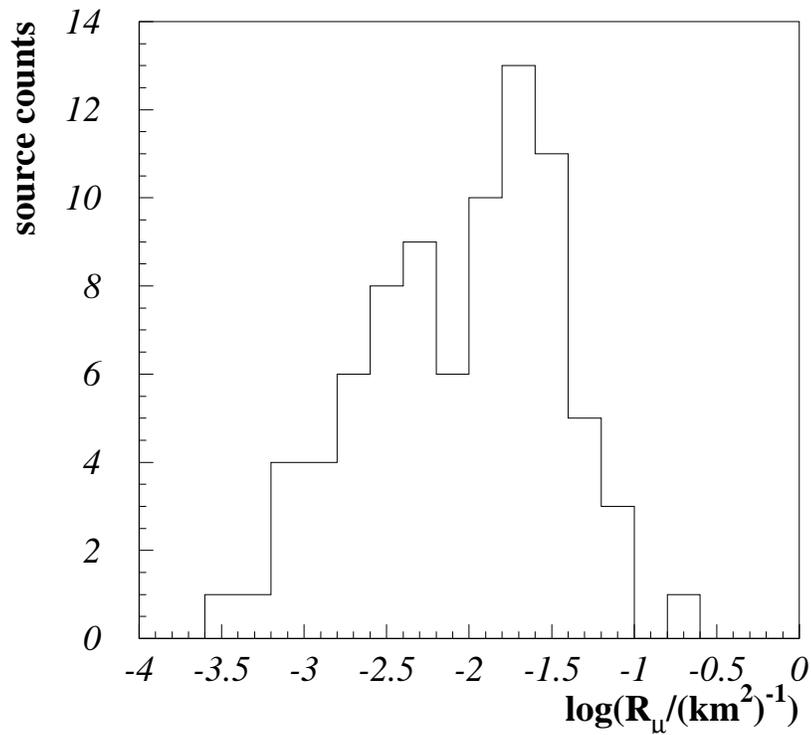}
}
\caption{Rate distribution for the $82$ bursts in the AMANDA subsample.}
\label{rates_ama}
\end{figure}
\clearpage
\bibliographystyle{elsart-num}


\begin{thebibliography}{10}
\expandafter\ifx\csname url\endcsname\relax
  \def\url#1{\texttt{#1}}\fi
\expandafter\ifx\csname urlprefix\endcsname\relax\def\urlprefix{URL }\fi

\bibitem{band}
D.~{Band}, et~al., Astrophys.~J. 413 (1993) 281.

\bibitem{hh_02}
F.~{Halzen}, D.~{Hooper}, Reports of Progress in Physics 65 (2002) 1025.

\bibitem{zhang}
B.~{Zhang}, P.~{Meszaros}, Intern.~Journ.~of Modern Phys.~A 19 (2004) 2385,
  astro-ph/0311321.

\bibitem{piran}
T.~{Piran}, Reviews of Modern Physics 76 (2004) 1143, astro-ph/0405503.

\bibitem{dar_rujula}
A.~{Dar}, A.~{de R{\'u}jula}, Physics Reports 405 (2004) 203.

\bibitem{wax_equi}
E.~{Waxman}, S.~R. {Kulkarni}, D.~A. {Frail}, Astrophys.~J. 497 (1998) 288.

\bibitem{guetta}
D.~{Guetta}, et~al., Astroparticle Physics 20 (2004) 429.

\bibitem{ppb}
{Particle Data Group}, {Phys.~Let.~B} 592 (2004) 1, {Particle Physics Booklet}.
\newline\urlprefix\url{http://pdg.lbl.gov}

\bibitem{WB}
E.~{Waxman}, J.~{Bahcall}, Phys.~Rev.~D 59 (1999) 23002.

\bibitem{050509b}
J.~{Hjorth}, et~al., Astrophys.~J.~Let. 630 (2005) L117.

\bibitem{050709}
J.~S. {Villasenor}, et~al., {Discovery of the short {$\gamma$}-ray burst GRB
  050709}, Nature 437 (2005) 855.

\bibitem{050724}
J.~{Gorosabel}, et~al., ArXiv Astrophysics e-prints.

\bibitem{band_lag}
{GSFC Lag-Luminosity Database},
  {http://heasarc.gsfc.nasa.gov/docs/cgro/analysis/lags/index.html} (2005).

\bibitem{norris}
J.~P. {Norris}, Astrophys.~J. 579 (2002) 386.

\bibitem{bonnell}
J.~T. {Bonnell}, et~al., Astrophys.~J. 490 (1997) 79.

\bibitem{mike}
M.~{Stamatikos}, {AMANDA Collaboration}, in: AIP Conf. Proc. 727: Gamma-Ray
  Bursts: 30 Years of Discovery, 2004, p. 146.

\bibitem{mike_in_prep}
M.~{Stamatikos}, et~al., In Prep. (2005).

\bibitem{mike_icrc05}
M.~{Stamatikos}, et~al., International Conference of Cosmic Rays, Pune (India),
  astro-ph/0510336 (2005).

\bibitem{pdflib}
CERN, PDFLIB - User's Manual, 8th Edition.

\bibitem{grv}
M.~{Gl\"uck}, E.~{Reya}, A.~{Vogt}, Z.~Phys. C~(53).

\bibitem{earthmodel}
A.~M. {Dziewonski}, D.~L. {Andersson}, Physics of the Earth and Planetary
  Interiors 25~(4) (1981) 297.

\bibitem{gaisser}
T.~K. Gaisser, Cosmic Rays and Particle Physics, Cambridge University Press,
  1990.

\end{thebibliography}
\end{document}